\documentclass[preprint,12pt]{elsarticle}

\usepackage{amssymb}

\usepackage{enumitem}
\usepackage[hyphens]{url}
\usepackage[hidelinks]{hyperref}
\usepackage{xcolor}
\usepackage{xspace}
\usepackage{colortbl}
\usepackage{caption}
\usepackage{hhline}
\usepackage{multirow}
\usepackage{ragged2e}
\usepackage{caption}
\usepackage{amsmath}
\usepackage{cleveref}
\newcolumntype{P}[1]{>{\RaggedRight\hspace{0pt}}p{#1}}
\usepackage{fontawesome}
\usepackage{pdflscape}
\usepackage{afterpage}
\usepackage{geometry}
\usepackage[framemethod=tikz]{mdframed}
\usepackage{subcaption}
\usepackage[most]{tcolorbox}
\hypersetup{breaklinks=true}
\usepackage{etoolbox}
\usepackage{hyphenat}

\makeatletter
\def\ps@pprintTitle{%
 \let\@oddhead\@empty
 \let\@evenhead\@empty
 \def\@oddfoot{}%
 \let\@evenfoot\@oddfoot}
 \makeatother

\mdfdefinestyle{mpdframe}{
    frametitlebackgroundcolor   =black!15,
    frametitlerule              =true,
    roundcorner                 =5pt,
    middlelinewidth             =1pt,
    innermargin                 =0.3cm,
    outermargin                 =0.3cm,
    innerleftmargin             =0.3cm,
    innerrightmargin            =0.3cm,
    innertopmargin              =0.5cm,
    innerbottommargin           =0.5cm
}

\journal{Journal of Systems and Software}

\newif\ifdraft
\drafttrue

\begin{document}
\ifdraft
  \newcommand{\jon}[1]{{\color{blue}\emph{Jon: #1}}\xspace}
  \newcommand{\arif}[1]{{\color{orange}\emph{Arif: #1}}\xspace}
  \newcommand{\gillian}[1]{{\color{magenta}\emph{Gillian: #1}}\xspace}
  \newcommand{\waqar}[1]{{\color{green}\emph{Waqar: #1}}\xspace}
  \newcommand{\mojtaba}[1]{{\color{cyan}\emph{Mojtaba: #1}}\xspace}
  \newcommand{\harsha}[1]{{\color{brown}\emph{Harsha: #1}}\xspace}
  \newcommand{\rifat}[1]{{\color{red}\emph{Rifat: #1}}\xspace}
\else
  \usepackage[disable]{todonotes}
  \newcommand{\jon}[1]{}
  \newcommand{\arif}[1]{}
  \newcommand{\gillian}[1]{}
  \newcommand{\waqar}[1]{}
  \newcommand{\mojtaba}[1]{}
  \newcommand{\harsha}[1]{}
  \newcommand{\rifat}[1]{}
\fi

\begin{frontmatter}



\title{Investigating End-Users' Values in Agriculture Mobile Applications Development: An Empirical Study on Bangladeshi Female Farmers}



\author[monash]{Rifat Ara Shams\corref{cor1}}
\ead{rifat.shams@monash.edu}
\author[rmit]{Mojtaba Shahin}
\ead{Mojtaba.Shahin@rmit.edu.au}
\author[monash]{Gillian Oliver}
\ead{Gillian.Oliver@monash.edu}
\author[data61]{Harsha Perera}
\ead{Harsha.Perera@data61.csiro.au}
\author[data61]{Jon Whittle}
\ead{Jon.Whittle@data61.csiro.au}
\author[monash]{Arif Nurwidyantoro}
\ead{Arif.Nurwidyantoro@monash.edu}
\author[data61]{Waqar Hussain}
\ead{Waqar.Hussain@data61.csiro.au}

\cortext[cor1]{Corresponding author}
\address[monash]{Department of Software Systems and Cybersecurity, Faculty of Information Technology, Monash University, 3800, Clayton, Australia}
\address[rmit]{School of Computing Technologies, RMIT University, 3000, Melbourne, Australia}
\address[data61]{CSIRO's Data61, 3168, Clayton, Australia}

\begin{abstract}
The omnipresent nature of mobile applications (apps) in all aspects of daily lives raises the necessity of reflecting end-users' values (e.g., \textit{fairness}, \textit{honesty}, \textit{social recognition}, etc.) in apps. However, there are limited considerations of end-users' values in apps development. Value violations by apps have been reported in the media and are responsible for end-users' dissatisfaction and negative socio-economic consequences. Value violations may bring more severe and lasting problems for marginalized and vulnerable end-users of apps, which have been explored less (if at all) in the software engineering community. One of the main reasons behind value violations is the lack of understanding of human values due to their ill-defined, ambiguous, and implicit nature. Therefore, understanding the values of the end-users of apps is the essential first step towards values-based apps development. This research aims to fill this gap by investigating the human values of Bangladeshi female farmers as a marginalized and vulnerable group of end-users of Bangladeshi agriculture apps. We conducted an empirical study that collected and analyzed data from a survey with 193 Bangladeshi female farmers, the end-users of Bangladeshi agriculture apps, to explore the underlying factor structure of Bangladeshi female farmers' values and the significance of demographics on their values. The results identified three underlying factors of Bangladeshi female farmers. The first factor comprises of five values: \textit{benevolence}, \textit{security}, \textit{conformity}, \textit{universalism}, and \textit{tradition}. The second factor consists of two values: \textit{self-direction} and \textit{stimulation}. The third factor includes three values: \textit{power}, \textit{achievement}, and \textit{hedonism}. We also identified strong influences of demographics on some of the values of Bangladeshi female farmers. For example, area has significant impacts on three values: \textit{hedonism}, \textit{achievement}, and \textit{tradition}. Similarly, there are also strong influences of household income on \textit{power} and \textit{security}. The results provide a direction for app developers on which values they should consider while developing agriculture apps for Bangladeshi female farmers.
\end{abstract}

\begin{keyword}
Human Values \sep Mobile Applications \sep Bangladeshi Female Farmers  \sep Empirical Study
\end{keyword}







\end{frontmatter}



\section{Introduction}
\label{sec:introduction}
Software is ubiquitous in daily lives which raises the necessity to consider end-users' human values such as \textit{transparency}, \textit{accessibility}, \textit{freedom}, \textit{independence}, \textit{fairness}, and \textit{tradition} in software. Human values are defined as ``desirable, trans-situational goals, varying in importance, that serve as guiding principles in people’s lives'' \cite{schwartz2013value}. Some recent studies focused on human values in software engineering (SE). For example, Mougouei et al. proposed a research roadmap to consider human values in software \cite{mougouei2018operationalizing}. Obie et al. proposed an automated technique using a natural language processing approach to detect values violations reported in app reviews \cite{obie2021first}. Another research conducted two case studies to identify the practices and challenges of addressing human values in SE \cite{hussain2020human}. A recent study proposed five intervention points in the Scaled Agile Framework (SAFe) where human values can be addressed \cite{hussain2022can}. In addition, some techniques have been proposed to consider human values in software development. They are Value-Based Requirements Engineering (VBRE) \cite{thew2018value}, Value-Sensitive Design (VSD) \cite{davis2015value}, Value-Sensitive Software Development (VSSD) \cite{aldewereld2015design}, Values-First SE \cite{ferrario2016values}, and Values Q-sort \cite{winter2018measuring}.

Despite these efforts, SE research and practices have limited considerations of human values in software \cite{mougouei2018operationalizing}. A recent study shows that only 16\% of 1350 papers published between 2015 and 2018 in top-tier SE journals and conferences were directly relevant to human values \cite{perera2020study}. There are also many examples of value violations in software. For example, Facebook allowed Cambridge Analytica to use 50 million Facebook users' personal data to gain a political advantage without the consent of the users \cite{Cadwalladr:2018}. Therefore, this incident violated Facebook users' values such as \textit{privacy} and \textit{trust}. Similarly, Instagram was blamed for the suicide of a British teenager because the recommendation algorithm of Instagram recommended self-harming images like any other interests such as sports or music in her newsfeed \cite{Crawford:2019}. This incident arguably violated society's values such as \textit{preservation of life}. Value violations in software are more destructive if the end-users are vulnerable and marginalized women in conservative societies. They usually face many social and cultural challenges which might be amplified because of value violations in the software they use \cite{sultana2018design}. For example, recent value violations occurred in 61\% (22 out of 36 apps) of menstruation apps, where they shared users' incredibly personal details with Facebook without the users' consent \cite{Nobody2019, al2020we}. This \textit{privacy} breach is a threat to the women's mental health and might have destructive impacts on their families and social lives. Therefore, addressing end-users' values in software is essential. It is more important for mobile software applications (apps) as with the growing number of smartphones, many organizations have been providing their products through apps. Apps are rich and complex \cite{stuurman2014design}, causing more usability and utility problems \cite{longoria2001designing} than other software. Therefore, addressing end-users' values in apps is also different from other software. However, there is little research on human values in apps from the perspective of vulnerable and underrepresented end-users (e.g., women in developing countries) \cite{shams2021measuring}.

It is argued that the essential first step towards developing end-users' values-based apps is understanding the values of the end-users \cite{shams2021measuring}. However, because of the ill-defined nature of human values, they are difficult to capture and express \cite{perera2020study}. As a result, there is a lack of understanding of human values \cite{perera2020study}. To address this issue, we conducted an empirical study that explored the values of Bangladeshi female farmers, vulnerable and underrepresented end-users of Bangladeshi agriculture apps. This study collected and analyzed data from a survey with 193 Bangladeshi female farmers. In one of our previous studies, we identified the value priorities of Bangladeshi female farmers and explored how their value priorities differ demographically \cite{shams2021measuring}. In this study, we investigated the possible underlying factor structure of Bangladeshi female farmers' values to explore whether there are different groups of Bangladeshi female farmers with similar values. We also investigated the significance of the demographics on their values. Our findings indicate:

\begin{itemize}
    \item Three (3) underlying factors of Bangladeshi female farmers' values. Factor1 consists of five values, Factor2 consists of two values, and Factor3 includes three values. For example, Factor2 comprises of \textit{self-direction} and \textit{stimulation}.
    \item Strong influences of two demographic groups (area and household income) on five values: \textit{hedonism}, \textit{achievement}, \textit{tradition}, \textit{power}, and \textit{security}.
\end{itemize}

\textbf{Paper Organization:} Section \ref{sec:background} describes the background of this research and the related work. Section \ref{sec:methodology}, briefly explains our research method and Section \ref{sec:results} reports the findings of this study. We discuss the findings in Section \ref{sec:discussion}. Section \ref{sec:ttv} discusses the possible threats to validity of this research. Finally, the research is concluded with possible future research directions in Section \ref{sec:conclusion}.
\section{Background and Related Work}
\label{sec:background}

\subsection{Human Values and Values Theory}

Social scientists have been conducting research to define and conceptualize human values since 1931 \cite{inbook}. Rokeach defined Human values as \textit{``a belief that a particular way of doing something is personally or socially preferable to the opposite ways''} \cite{rokeach1973nature}. Later, Shalom H. Schwartz gave a simpler definition of human values as \textit{``things that people hold important in their life''} \cite{schwartz2012overview}. A recent work has summarized seven different definitions of human values as \textit{``guiding principles of what people consider important in life''} \cite{cheng2010developing}. Values are intertwined with feelings \cite{schwartz2012overview}, and therefore, respect for values would bring enjoyment for an individual while relative importance for each value may differ for each person \cite{rokeach1973nature,schwartz2012overview,Schwartz2017}. Moreover, values have been used to identify personality and to understand what is and is not important in life \cite{allport1960study}. 

Social scientists have been conducting research to conceptualize human values since 1931 \cite{inbook}. Allport and Vernon proposed the first values theory in 1931 on personality typology \cite{vernon1931test}. Expanding their theory, Allport et al. classified values into six categories: theoretical, economic, aesthetic, social, political, and religious \cite{allport1960study} in 1960. A few years later, Rokeach conducted a series of surveys with people from different ethnic and social backgrounds and proposed 36 values, where he claimed them as universal values \cite{rokeach1973nature}. In 1980, Hofstede introduced two values categories, namely, desired and desirable i.e., what people actually desire and what they think ought to be desired respectively \cite{hofstede1980culture}. In 1992, Shalom H. Schwartz proposed a values theory that identified ten (10) human values defined by their motivational goals and measured from 58 value items \cite{schwartz1992universals}. In 2004, Parashar et al. divided values into two concepts, namely, individual behavior (micro values) and cultural practices (macro values) \cite{parashar2004perception}. In 2008, Inglehart identified two dimensions of values according to the cross-cultural variations as post-materialist and self-expression values \cite{inglehart2008changing}. In 2010, a meta-inventory of human values was introduced by analyzing 12 value inventories \cite{cheng2010developing}. The most recent values model was introduced in 2014 by Gouveia et al., in which six basic value categories were identified (existence values, promotion values, normative values, suprapersonal values, excitement values, and interactive values) \cite{gouveia2014functional}.

\begin{figure}[htb]
        \centering
        \includegraphics[width=0.75\textwidth]{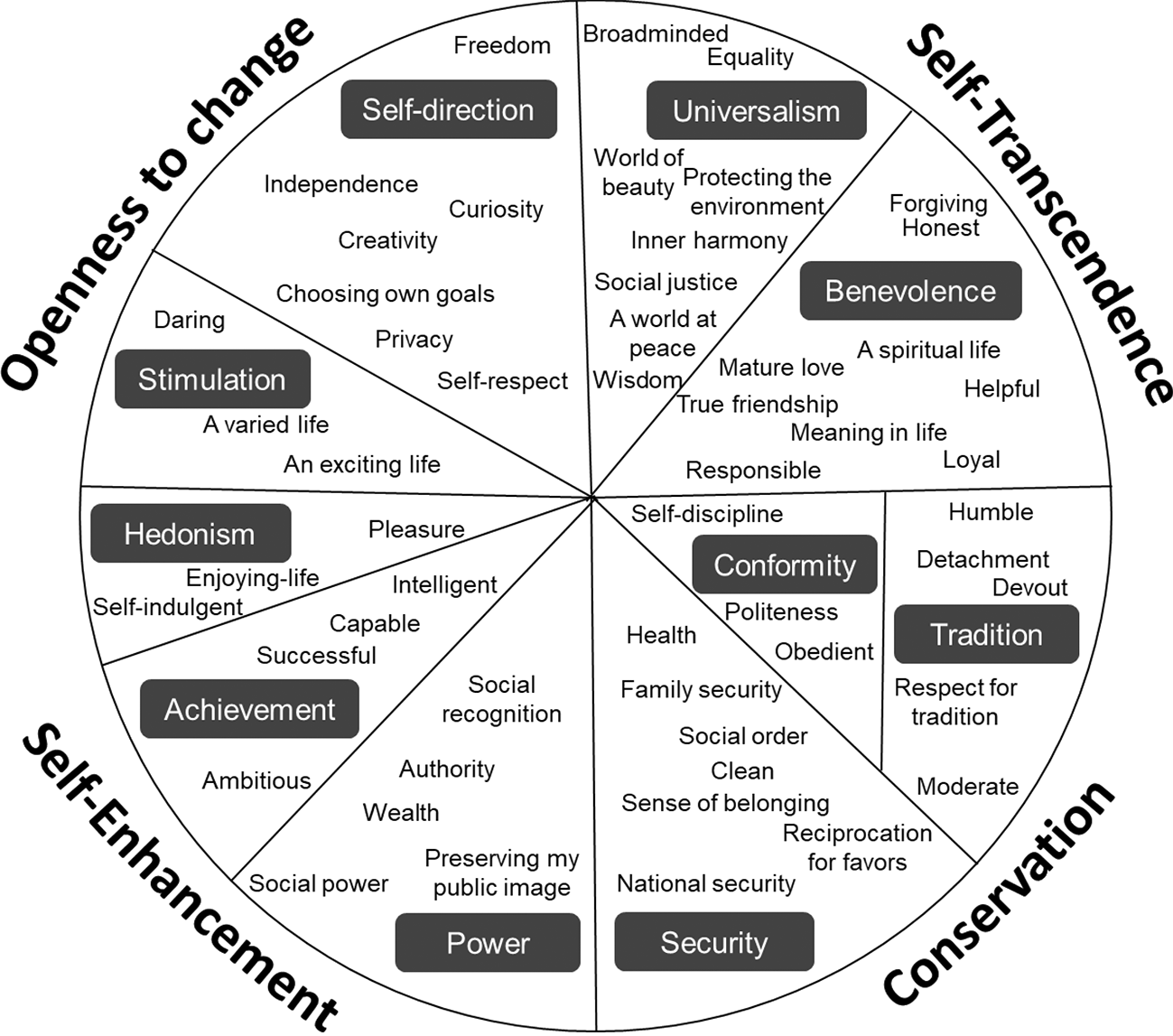}
        \caption{Schwartz's theoretical model of basic human values \cite{vcaic2019value}}
        \label{fig:SchwartzTheory}
    \end{figure}

In this study, we used Schwartz's theory of basic human values to explore Bangladeshi female farmers' values. Schwartz's values theory is recognised as the most cited, most comprehensive and widely adopted values theory to date \cite{thew2018value, ferrario2014software} in social science along with other areas such as computer science \cite{barcelo2014social} and software engineering \cite{ferrario2014software}. Moreover, this theory has been tested in different settings with variations according to geography, culture, language, religion, politics, age, education, and gender \cite{schwartz2001extending, schwartz1992universals}. In this theory, Schwartz divided values into ten main categories (self-direction, stimulation, hedonism, achievement, power, security, conformity, tradition, benevolence, and universalism) based on their motivational goals. These main value categories are measured from 58 value items. Values located close to each other are congruent and those further apart are opposite in nature \cite{schwartz1992universals, schwartz2012overview}. \autoref{fig:SchwartzTheory} shows Schwartz's theoretical model of basic human values of ten main value categories with the corresponding value items to measure those main value categories \cite{schwartz1992universals}.

\subsection{Value Measurement Instrument}
Schwartz proposed the following two types of instruments to measure human values suitable for all cross-cultural settings.
        
    \subsubsection{Schwartz Value Survey (SVS)}
    The first instrument to measure human values developed by Schwartz is Schwartz Value Survey (SVS) \cite{schwartz2012overview}. Based on the basic values theory of Schwartz, this survey presents two lists with 30 value items and 27 value items respectively (see the 57 SVS items in \cite{kusurkar2015critical}) to the participants to describe potentially desirable end-states and potentially desirable ways of acting \cite{schwartz2012overview}. Each value item demonstrates the motivational goal of the corresponding value. Each item contains an explanatory statement in parentheses that clarifies the meaning of that value item \cite{schwartz2012overview}. For example, the item, \textit{`equality (equal opportunity for all)'} refers to the main value \textit{universalism}. Similarly, \textit{`authority (the right to lead or command)'} is a value item of \textit{power}. The participants are then asked to respond to each value item as a guiding principle in their lives on a 9-point scale: 7 (of supreme importance), 6 (very important), 5 (unlabeled), 4 (unlabeled), 3 (important), 2 (unlabeled), 1 (unlabeled), 0 (not important), -1 (opposed to my values). The score of each value item is calculated by the average rating of that item given by all the participants \cite{schwartz2012overview}.
    
    
    \subsubsection{Portrait Values Questionnaire (PVQ)}
    According to Schwartz, the alternative instrument to measure human values is Portrait Values Questionnaire (PVQ). In PVQ, there are short text descriptions (portraits) of characteristics that reflect a person's goals, aspirations, and wishes leading to a particular value \cite{schwartz2012overview}. For example, the portrait, \textit{``It's very important to her to help the people around her. She wants to care for their well-being''} describes a person for whom \textit{`benevolence'} is an important value. Based on \textit{``How much like you is this person?''}, the participants are asked to rate each portrait on a 6-point scale: 6 (Very much like me), 5 (Like me), 4 (Somewhat like me), 3 (A little like me), 2 (Not like me) and 1 (Not like me at all) \cite{schwartz2001extending}.
    
    PVQ is easy to understand as it is less abstract, more concrete, more context-bound, and less complicated than the SVS \cite{schwartz2001extending}. Therefore, PVQ is suitable for less-educated people \cite{beierlein2012testing} like Bangladeshi female farmers. 68\% of Bangladeshi farmers' education level is less than Year 10, and 25\% have no education at all \cite{islam2011factors}. Furthermore, the female literacy rate is lower than the male in Bangladesh \cite{ferdaush2011gender}. Therefore, we selected PVQ as the value measurement instrument to explore the values of Bangladeshi female farmers.

    There are several versions of PVQ. In 2001, Schwartz introduced 40 items PVQ (PVQ-40) \cite{schwartz2003proposal}, which has been shortened to 21 items (PVQ-21) in 2002 to include in the semiannual European Social Survey (ESS) \cite{cieciuch2012comparison, jacques2016towards}. Schwartz also developed 20 items and 29 items PVQ \cite{schwartz2003proposal}. Furthermore, Bubeck and Bilsky developed 29 items PVQ in 2002 \cite{bubeck2004value}, and Knoppen and Saris developed 33 items PVQ in 2009 \cite{beierlein2012testing}. This research used 40 items PVQ because it is more suitable for cross-cultural research \cite{cieciuch2012comparison}. It is also argued that 40 items PVQ may lead to more accurate results as it is more elaborate and contains more items in each value \cite{cieciuch2012comparison}.
    

\subsection{Advanced Statistical Methods to Elicit Human Values}
Advanced statistical methods such as factor analysis are used in a few research to elicit human values. For example, Schwartz applied factor analysis to elicit the ten main value categories distinguished by their motivational goals \cite{schwartz1994there}. Later, Schwartz et al. confirmed his proposed ten main value categories by applying confirmatory factor analysis on 10,857 samples from 27 countries \cite{schwartz2004evaluating}. Another study applied factor analysis on a survey data to explore the values of Chinese students around the world and their relations with culture \cite{chinese1987chinese}. Braithwaite and Law adopted factor analysis on the collected dataset to examine the adequacy of Rokeach value survey in terms of comprehensive coverage of the value domain \cite{braithwaite1985structure}. Allen conducted regression and factor analysis to investigate the influences of human values on consumer purchase decisions \cite{allen2001practical}. Therefore, advanced statistical analysis can be considered as a comprehensive approach to elicit human values which we used to investigate the factor structure of Bangladeshi female farmers' values and the significance of demographics on their values.
    

\subsection{Prevalence of Human Values in Software Engineering}
Recently, a few studies have focused on human values in software engineering (SE) contexts. For example, the importance of considering human values in software engineering was described by Mougouei et al, where they also identified the research gap of eliciting human values \cite{mougouei2018operationalizing}. Ferrario et al. proposed a framework to consider values during SE decision making processes \cite{ferrario2016values}. Another study investigated values at the system, personal, and instantiation levels of SE by proposing a value measurement tool named Values Q-Sort \cite{winter2018measuring}. A recent study introduced a framework to integrate, trace, and evaluate human values throughout the software development life cycle (SDLC) \cite{perera2020continual}. Thew and Sutcliffe complemented the existing analysis of non-functional requirements by eliciting stakeholders' values through their proposed approach named Values Based Requirement Engineering (VBRE) \cite{thew2018value}. Another study introduced a framework values-based implications of software design patterns \cite{hussain2018integrating}. Aldewereld et al. translated values into more concrete features during software development by introducing a design for values approach named Value-Sensitive Software Development (VSSD) \cite{aldewereld2015design}.

Despite the emerging concerns of human values in software engineering, only 16\% papers of top-tier SE journals and conferences from 2015-2018 were directly relevant to human values. Furthermore, in one of our studies with Bangladeshi agriculture apps, we identified almost 50\% of the end-users' values are ignored/violated in those apps \cite{shams2020society}. Therefore, we argue that there is a room for development in addressing human values in software engineering. We also argue that eliciting end-users' values should be the first step towards this goal as eliciting human values is necessary to develop value requirements for the Software Development Life Cycle (SDLC) \cite{shams2021measuring}.


\subsection{Bangladesh, Women in Agriculture and Smartphones}
Bangladesh is an agriculture-dependent country due to the geographical setting \cite{chowhan2020role}. In Bangladesh, agriculture contributes 14.23\% of GDP in 2019 \cite{chowhan2020role}. Agriculture is the only source of earning for more than 80\% people in Bangladesh \cite{faroque2021effect}. 38\% of the labor force are involved in the agricultural sector (2019) \cite{WorldBankEmployment2021}. Nowadays, women are also participating in agricultural activities, resulting in the ``\textit{feminization of agriculture}'' \cite{jaim2011women}. Bangladeshi women, less than 20\% of the agricultural labour force, started participating in agricultural activities in 1999/2000 \cite{sraboni2014empowered}. Their participation was increased to 33.6\% in 2010 \cite{sraboni2014empowered}, more than 50\% in 2016 \cite{FAO2016}, and 58\% in 2019 \cite{WorldBankFemaleAgri2021}.

Mobile phones play an important role in agriculture as they provide the opportunity for farmers to access information regarding agricultural initiatives \cite{stillman2020after}. According to the Bangladesh Telecommunication Regulatory Commission (BTRC), there are 181.53 million mobile phone subscribers in Bangladesh till November 2021 \cite{BTRC2021}. As the population of Bangladesh is more than 164 million (2020) \cite{WorldBankBDPopulation2020}, it can be assumed that almost all people use mobile phones and some people use more than one mobile phone. The statistics by BTRC in October 2021 also show that the total number of Internet subscribers in Bangladesh is 129.18 million \cite{BTRCInternet2021}. Moreover, there is a potential future prospect of the smartphones market in Bangladesh as the number of smartphone users has been radically increasing in Bangladesh because of the availability and low price of smartphones \cite{deb2019investigation, ahmed2018acceptance}. As smartphones are easily accessible, the majority of Internet subscribers use the Internet via smartphones \cite{karim2010digital}. With the increase of the use of the Internet, mobile apps have become increasingly popular.

Agriculture mobile apps play an important role in agriculture development, resulting the popularity of agriculture apps in recent years. 35 Bangladeshi agriculture apps are available in Google play (2018) \cite{shams2020society} which are used by the stakeholders in agriculture, including farmers \cite{rahman2020utility, shams2021measuring}. Bangladeshi farmers used these apps to know about weather forecast which help them taking precautions \cite{chowhan2020role}. In addition, information such as agricultural problem solving, detecting diseases, finding disease solutions, and using the recommended medicines \cite{roshidul2016potential, kundu2017smart} are also available in those apps which are useful for the farmers.

There are a few studies on Bangladeshi agriculture apps. For example, Chowhan and Ghosh explained the role of mobile apps on Bangladesh agriculture and its future scope \cite{chowhan2020role}. Sharma conducted a case study of agriculture apps in Bangladesh and India to explore the importance of agriculture mobile apps in farmers' empowerment \cite{sharma2019ikhedut}. Another study identified the techniques suitable to develop agriculture apps in local languages to identify diseases and management of maize crop in Bangladesh \cite{roshidul2016potential}. However, we did not identify any published research on Bangladeshi farmers' values in agriculture apps other than our works. One of our papers identified the value priorities of Bangladeshi female farmers \cite{shams2021measuring}. In our another paper, the present and missing values of the users of Bangladeshi agriculture apps were explored by analyzing user reviews \cite{shams2020society}. Our other study determined the extent to which the existing Bangladeshi agriculture apps reflect Bangladeshi female farmers' values and identified the strategies to address their values in apps from both the end-users' and app practitioners' perspectives \cite{shams2021human}. In this research, we explored the possible underlying factor structure of Bangladeshi female farmers' values and the significance of the demographics on their values.
\section{Methodology}
\label{sec:methodology}

This research is funded by an Information and Communication Technology for Development (ICT4D) project named PROTIC (Participatory Research and Ownership with Technology, Information and Change) of a multinational charitable organization, Oxfam Bangladesh, in collaboration with Monash university. PROTIC works for the resilience of Bangladeshi female farmers through the use of ICT \cite{stillman2020after}. This study aims to gain a comprehensive understanding of the values of Bangladeshi female farmers to help apps developers develop Bangladeshi female farmers' values-based agriculture mobile apps. With this goal, we formulated the following two research questions. 

   \noindent \textit{\textbf{RQ1.} What is the possible underlying factor structure of Bangladeshi female farmers' values?}
   
   \noindent \textit{\textbf{RQ2.} How much influence do the demographics have on the values of Bangladeshi female farmers?}
    
    \begin{figure*}[!htbp]
            \centering
            \includegraphics[width=1\textwidth]{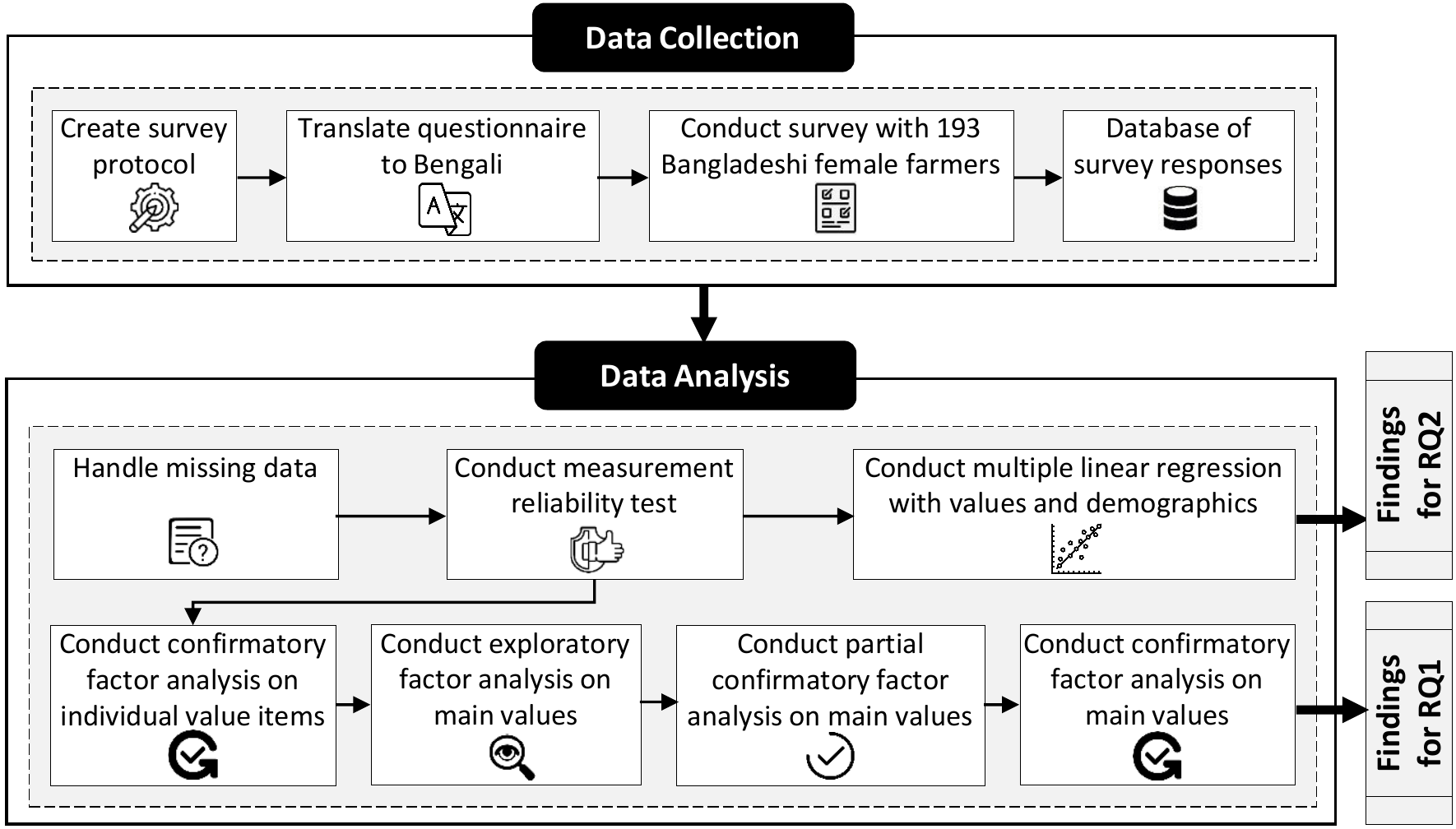}
            \caption{An overview of the research method}
            \label{fig:research_method}
    \end{figure*}
   
To answer the research questions, we conducted an empirical study where we collected data through a survey with 193 Bangladeshi female farmers. The survey data was previously used to identify the value priorities of Bangladeshi female farmers and how their value priorities differ demographically by applying descriptive statistical methods on the data \cite{shams2021measuring}. This research used the same survey data from different perspectives to answer the research questions using advanced statistical analysis (factor analysis and multiple regression analysis). \autoref{fig:research_method} shows an overview of the research method of this study. Before we conducted this study, ethics approval was acquired from Monash university Human Research Ethics Committee (MUHREC) on 21/11/2019.
    
    
\subsection{Protocol}
\label{subsec:survey}
\begin{figure}[!htbp]
        \centering
        \centering \includegraphics[width=0.6\textwidth]{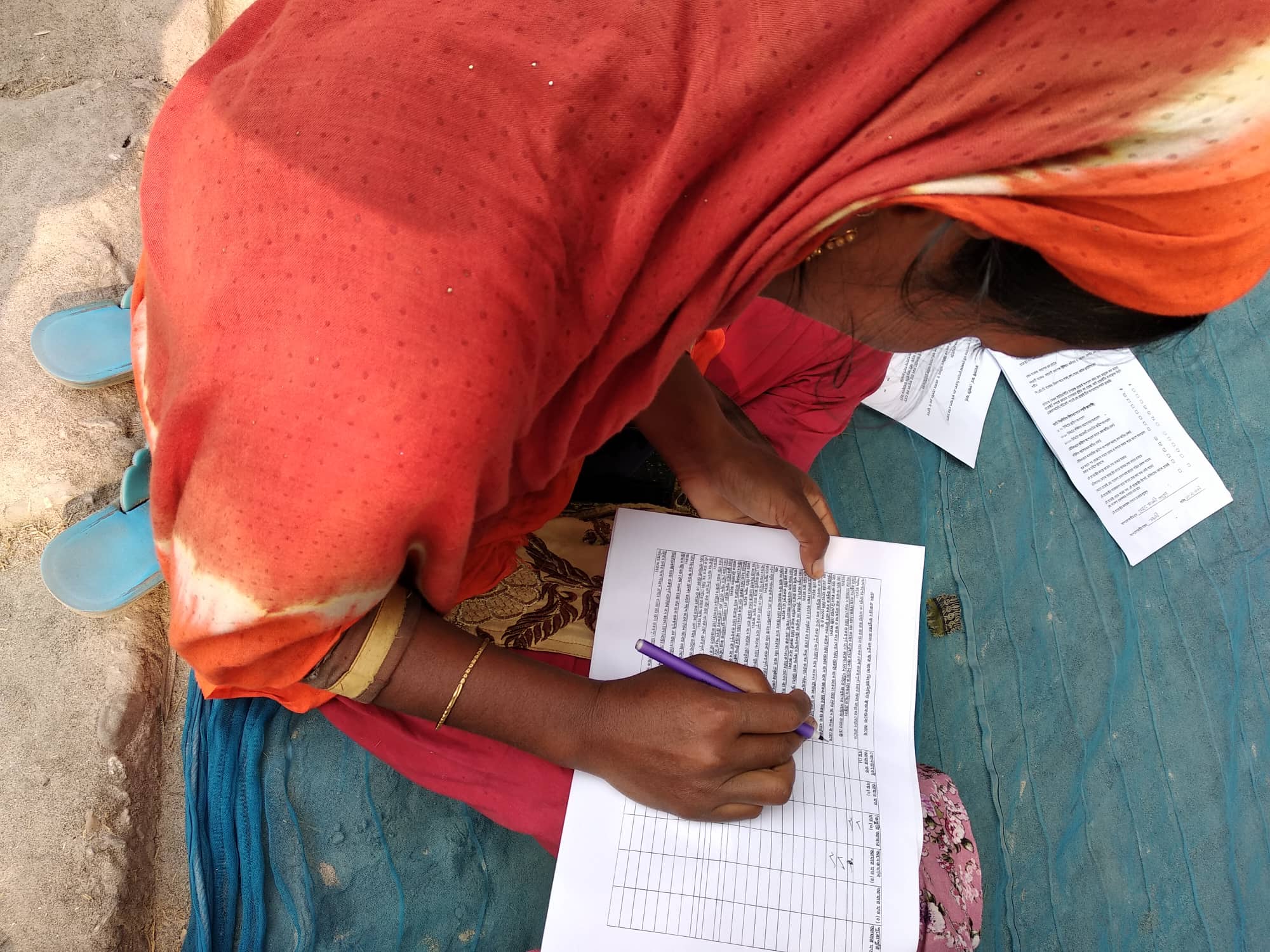}
        \caption{Conducting the survey in natural setting}
        \label{fig:survey_FG_conduction}
    \end{figure}

We conducted a paper-based survey with 193 Bangladeshi female farmers who use agriculture mobile apps in their daily agricultural activities. The survey was conducted in December 2019 at their villages. The participants felt comfortable as the survey was conducted in their natural setting. It took 9 days to complete the survey with 193 participants. \autoref{fig:survey_FG_conduction} shows a picture of conducting the survey.

The survey was divided into three parts. The first part worked as an ice-breaker session where we spent approximately 15 minutes having a friendly chat with the female farmers to put them at their ease. In the second part, the research was explained to the participants along with its objectives, expected outcomes, and confidential data storage. They were also provided with an explanatory statement of this research and the consent form. In this part, the participants were provided with proper guidance on how to complete the survey. The last part was the survey which took approximately 15 minutes to complete. 

In this survey, we used a value measurement tool named Portrait Values Questionnaire (PVQ) proposed by Schwartz \cite{schwartz2003proposal} to identify the values of the female farmers. Each portrait (i.e., item) in PVQ describes a person's characteristic that reflects the importance of a particular value. For example, a portrait, ``Thinking up new ideas and being creative is important to her. She likes to do things in her own original way" describes a person for whom \textit{Self-direction} is an important value. The participants were then asked to compare themselves with each portrait and rate on a 6-point Likert scale: 6 (Very much like me), 5 (Like me), 4 (Somewhat like me), 3 (A little like me), 2 (Not like me) and 1 (Not like me at all) \cite{schwartz2001extending}. In our survey, there were 40 portraits (i.e., items) to understand the values of Bangladeshi female farmers \cite{schwartz2003proposal} (see the survey questionnaire \cite{shams_rifat_ara_2022_6370062}). We also collected demographic data from the participants through six questions. For example, ``which area are you from?", ``which age group do you belong to?", ``what is the highest degree or level of school you have completed? If currently enrolled, highest degree received?" etc. (see the demographic questions \cite{shams_rifat_ara_2022_6370062}).

    \begin{figure}
        \centering \includegraphics[width=1\textwidth]{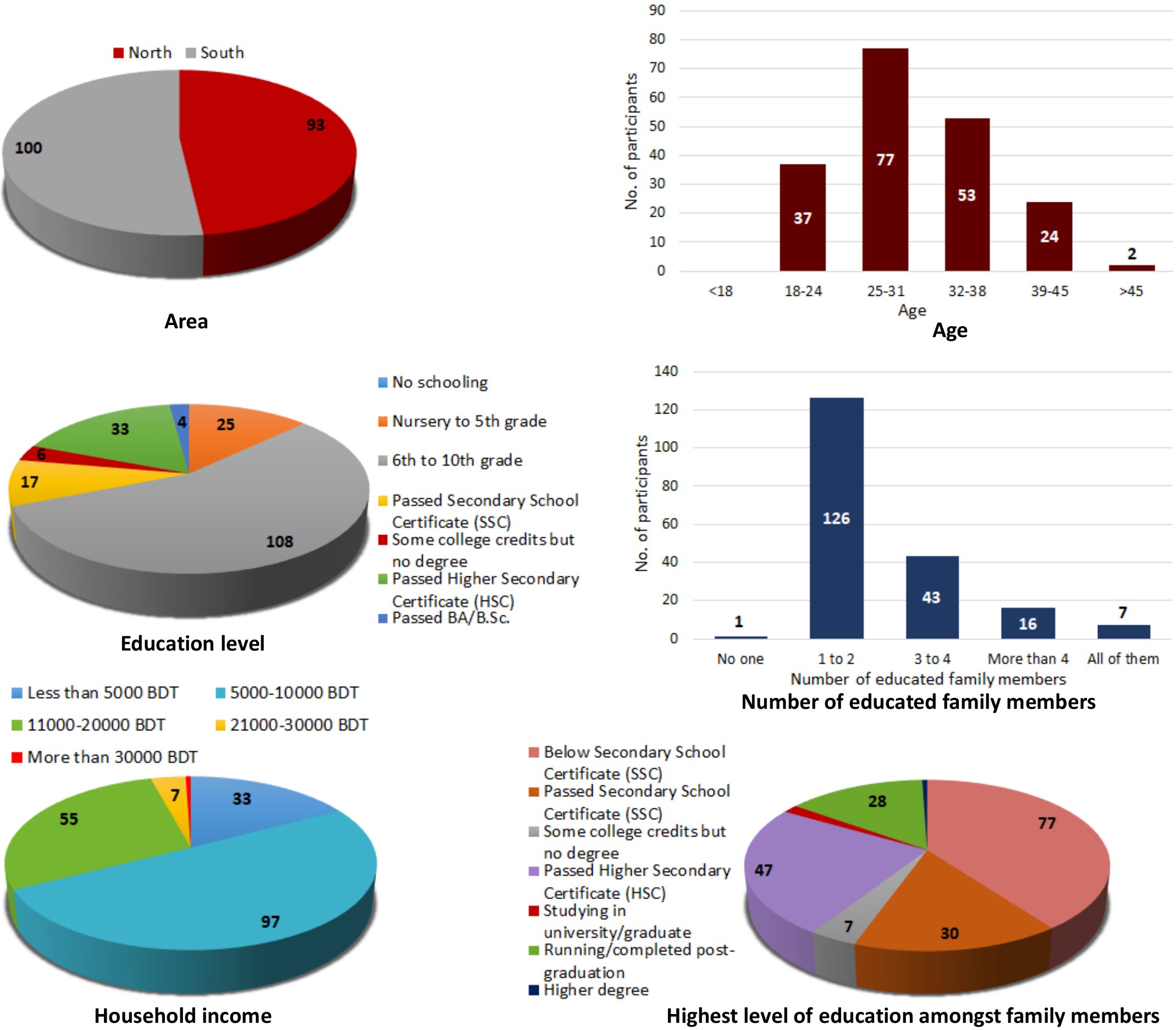}
        \caption{Demographics of the survey participants}
        \label{fig:survey_demographics}
    \end{figure}

Although the PVQ questionnaire has been translated in different languages (even in local languages such as Afrikaans \cite{schwartz2001extending}) and conducted in various countries \cite{romanyuk2017personality}, we found that the questionnaire has no Bengali version and PVQ was never conducted in Bangladesh. Therefore, we translated the PVQ questionnaire from English to Bengali. We involved an independent translator who had no previous knowledge of the PVQ questionnaire. At first, the PVQ questionnaire was translated from English to Bengali by the first author of this paper who is Bangladeshi. Then, the independent translator translated the Bengali version back to English. After that, we compared the back-translated version with the original version of PVQ and found some inconsistencies. Then, the first author translated the questionnaire to Bengali again to solve these issues. The independent translator again back-translated the Bengali version of PVQ to English. After comparing this version with the original one, we hardly found any inconsistencies. Therefore, we used this Bengali version of PVQ in our survey. The collected data were transferred to an Excel sheet and SPSS, stored in Google Drive for the data analysis. This file was shared with all the authors of this paper. We ensured the confidentiality of the paper-based data by storing it in a locked cabinet at our lab in Monash university.

\subsection{Participants}

The main criterion for selecting the survey participants was to find Bangladeshi female farmers who use agriculture apps in their daily farming activities. Oxfam Bangladesh helped us select the participants. Oxfam provided smartphones to 100 female farmers from the Northern part (Sandy area \cite{anik2012climate}) and 100 from the Southern part (Coastal area \cite{rakib2019investigation}) of Bangladesh, as well as trained them on how to use smartphones and different agriculture mobile apps. As our selection criterion was precisely similar to those 200 female farmers, we recruited them as the survey participants. Seven women were unavailable during the survey, so we collected data from 193 participants.

\textbf{Survey Participants Characteristics.} \autoref{fig:survey_demographics} shows the corresponding numbers of survey participants in each demographic group.


\subsection{Data Analysis}
\label{da}

We analyzed the survey data by following the seven steps as described below.

    \textbf{Step 1: Data Pre-processing.} At first, the paper-based survey data was transferred to SPSS software \cite{landau2004handbook}. We transferred the participants' responses as well as their demographic data but we did not transfer their names so that their identity can remain protected. 
    
    We also found missing responses for a few questions and some multiple responses for a particular question. We considered the multiple responses as missing responses \cite{vriens2006dealing} and after that, we found 275 (3.56\%) missing responses out of 7720 responses. We used multiple imputation technique in SPSS to replace the missing values with plausible responses \cite{zhong2018application}, as \textit{multiple imputation} technique reduces the risk of wrong estimations due to reduced sample size and increases the statistical power \cite{vriens2006dealing}.
    
    \textbf{Step 2: Measurement Reliability Test.} It was essential to examine the reliability of our value measurement tool, PVQ, because a measurement tool is not valid unless it is reliable \cite{tavakol2011making}. Measurement reliability is used to examine the consistency of responses to a set of items that emerge a concept \cite{shelby2011beyond}. Cronbach's alpha is one of the most widely used techniques to measure the reliability in the social and organizational sciences \cite{bonett2015cronbach}. Cronbach's alpha describes the extent to which the responses are correlated to each other \cite{shelby2011beyond}. We applied Cronbach's alpha using the SPSS software to measure the consistency of the responses of the items under each value category (e.g., consistency of the four responses under the value category of \textit{Benevolence}, six responses under \textit{Universalism}). The formula of Cronbach's alpha is \cite{cortina1993coefficient}:
    
    \[ N^2 (\dfrac{Mean(Cov)}{Sum(Var/Cov)}) \]
    
    Here, N is the number of items in the scale, Mean(Cov) is the mean of inter-item covariance, and Sum(Var/Cov) is the sum of all the elements in the variance-covariance matrix \cite{cortina1993coefficient}.
    
    \textbf{Step 3: Confirmatory Factor Analysis on Individual Value Items.} Confirmatory Factor Analysis (CFA) is a statistical method that deals with measurement models to explore the factors of a set of observed variables \cite{suhr2006exploratory, brown2012confirmatory}. In other words, CFA tests a hypothesis of the relationship between manifest variables (observed variables) and latent variables (unobserved variables) statistically \cite{suhr2006exploratory}. Therefore, we applied CFA on our survey data to check if the data are aligned with the hypothesis developed by Schwartz regarding which particular items are used to measure a value (e.g., 4 items were used to measure \textit{Benevolence}, 6 items to measure \textit{Universalism}). We used IBM SPSS AMOS 26 to conduct the CFA with the 40 items to test the relations between the manifest variables (e.g., 40 individual value items) and the latent variables or factors.
    
    \textbf{Step 4: Exploratory Factor Analysis on Main Value Categories.} Exploratory Factor Analysis (EFA) is a variable reduction statistical technique that explores the possible underlying factor structure of a set of observed variables \cite{suhr2006exploratory}. Therefore, to identify the possible factors from the 10 main value categories, we conducted EFA with principal component extraction and varimax rotation using IBM SPSS 26. At this stage, the 10 main value categories were calculated by averaging the items under each category (e.g., average of the 4 items to calculate \textit{Benevolence}, average of the 6 items to calculate \textit{Universalism}). Later, we applied the EFA on these 10 value categories which worked as observed variables during the EFA. The decision on how many factors should be considered depends on the eigenvalues of each factor \cite{larsen2010estimating}. According to the suggestions of Guttman, Kaiser's criterion retains factors with eigenvalue greater than 1 \cite{suhr2006exploratory}. Therefore, in this study, we also followed the criterion and considered the factors whose eigenvalues are greater than 1.
    
    \textbf{Step 5: Partial Confirmatory Factor Analysis on Main Value Categories.} Exploratory Factor Analysis (EFA) suggests that research should conduct Confirmatory Factor Analysis (CFA) to confirm the factors derived from the EFA \cite{gignac2009partial}. Before proceeding to CFA, it is a good practice to conduct Partial Confirmatory Factor Analysis (PCFA) which is a data reduction technique that lies between EFA and CFA to understand whether CFA is justifiable and whether the factors derived from the EFA have a strong chance of getting confirmed by the CFA \cite{gignac2009partial}. Therefore, we have conducted PCFA using IBM SPSS 26 and Excel to test if the EFA results are good fit for CFA. For this purpose, we have used four close-fit indexes \cite{gignac2009partial}: Normed Fit Index (NFI) \cite{bentler1980significance}, Comparative Fit Index (CFI) \cite{bentler1990comparative}, Tucker-Lewis Index (TLI) \cite{tucker1973reliability}, Root Mean Square Error of Approximation (RMSEA) \cite{browon1993alternative}.
    
    \[ NFI = \dfrac{(\chi^2_{Null} - \chi^2_{Implied})}{(\chi^2_{Null})} \]
    \[ TLI = \dfrac{(\chi^2_{Null}/df_{Null}) - (\chi^2_{Implied}/df_{Implied})}{[(\chi^2_{Null}/df_{Null}) - 1]} \]
    \[ CFI = 1 - \dfrac{(\chi^2_{Implied} - df_{Implied})}{(\chi^2_{Null} - df_{Null})} \]
    \[ RMSEA = \sqrt{\dfrac{\chi^2_{Implied} - df_{Implied}}{(N-1) * df_{Implied}}} \]
    
    Here, \( \chi^2_{Implied} \) is the Maximum Likelihood Estimation (MLE) chi-square associated with the residual correlation matrix, \(df_{Implied} \) is the degrees of freedom associated with the chi-square implied, N is the sample size, \( \chi^2_{Null} \) is the MLE null model chi-square, \(df_{Null} \) is the degrees of freedom associated with the null model chi-square \cite{gignac2009partial}.
    
    \textbf{Step 6: Confirmatory Factor Analysis on Main Value Categories.} To confirm the relations between the manifest variables (10 main value categories) and the latent variables or factors derived from the Exploratory Factor Analysis (EFA), we conducted Confirmatory Factor Analysis (CFA) on the 10 main value categories by using IBM SPSS AMOS 26.
    
    \textbf{Step 7: Multiple Linear Regression Analysis on Values with Demographics.} We conducted a multiple linear regression analysis to check the significance of demographics (area, age, education level, household income, number of educated family members, highest level of education among family members) on values (e.g., Benevolence, Universalism, Self-direction etc.). Multiple linear regression is a linear regression model which explores the relationship between one dependent variable and multiple independent variables \cite{yan2003linear}. Therefore, we conducted the multiple linear regression analysis ten times using IBM SPSS 26 to check the significance of the demographics (independent variables) on the ten main value categories (dependent variables).

\section{Results}
\label{sec:results}

This section presents the results of the survey with 193 Bangladeshi female farmers.

\subsection{RQ1: Factor Structure of Bangladeshi Female Farmers' Values}
\label{Results_RQ1}
After the data pre-processing, we applied the following five steps of statistical analysis on the survey data to explore the factor structure of Bangladeshi female farmers' values (RQ1). The results of these five steps are discussed below.

    \subsubsection{Measurement Reliability Test}
    \label{reliability_test}
    
    \begin{table*}
\centering
\caption{Reliability analysis of the latent variables for all survey respondents}
\label{table:cronbach_alpha}
\resizebox{0.88\textwidth}{!}{%
\renewcommand{\arraystretch}{1.15}
\begin{tabular}{|p{3.6cm}|P{4.1cm}|p{3.6cm}|p{4.3cm}|}
\hline
\textbf{Variables}               & \textbf{Cronbach's alpha} & \textbf{Attributes}                  & \textbf{Cronbach's alpha if item deleted} \\\hline
                                 &                                               & Benevolence1 &  0.70          \\
                                 &                                               & Benevolence2 &  0.68          \\
                                 &                                               & Benevolence3 &  0.71          \\
\multirow{-4}{*}{Benevolence}    & \multirow{-4}{*}{0.75}                        & Benevolence4 &  0.69          \\\hline
                                 &                                               & Universalism1                        &  0.61           \\
                                 &                                               & Universalism2                        &  0.64          \\
                                 &                                               & Universalism3                        &  0.66          \\
                                 &                                               & Universalism4                        &  0.68          \\
                                 &                                               & Universalism5                        &  0.66          \\
\multirow{-6}{*}{Universalism}   & \multirow{-6}{*}{0.69}                        & Universalism6                        &  0.66           \\\hline
                                 &                                               & Self-direction1                      &  0.62          \\
                                 &                                               & Self-direction2                      &  0.64          \\
                                 &                                               & Self-direction3                      &  0.66          \\
\multirow{-4}{*}{Self-direction} & \multirow{-4}{*}{0.70}                         & Self-direction4                      &  0.64         \\\hline
                                 &                                               & Stimulation1                         & 0.71                                                         \\
                                 &                                               & Stimulation2                         & 0.52                                                         \\
\multirow{-3}{*}{Stimulation}    & \multirow{-3}{*}{0.68}                        & Stimulation3                         & 0.49                                                         \\\hline
                                 &                                               & Hedonism1                            & 0.38                                                         \\
                                 &                                               & Hedonism2                            & 0.59                                                         \\
\multirow{-3}{*}{Hedonism}       & \multirow{-3}{*}{0.54}                        & Hedonism3                            & 0.31                                                         \\\hline
                                 &                                               & Achievement1                         &  0.43          \\
                                 &                                               & Achievement2                         &  0.58          \\
                                 &                                               & Achievement3                         &  0.53          \\
\multirow{-4}{*}{Achievement}    & \multirow{-4}{*}{0.61}                        & Achievement4                         &  0.58          \\\hline
                                 &                                               & Power1                               & 0.59                                                         \\
                                 &                                               & Power2                               & 0.41                                                         \\
\multirow{-3}{*}{Power}          & \multirow{-3}{*}{0.57}                        & Power3                               & 0.38                                                         \\\hline
                                 &                                               & Security1                            &  0.67          \\
                                 &                                               & Security2                            &  0.62          \\
                                 &                                               & Security3                            &  0.65           \\
                                 &                                               & Security4                            &  0.65          \\
\multirow{-5}{*}{Security}       & \multirow{-5}{*}{0.70}                         & Security5                            &  0.65           \\\hline
                                 &                                               & Conformity1                          &  0.65          \\
                                 &                                               & Conformity2                          &  0.51          \\
                                 &                                               & Conformity3                          &  0.56          \\
\multirow{-4}{*}{Conformity}     & \multirow{-4}{*}{0.65}                        & Conformity4                          &  0.60          \\\hline
                                 &                                               & Tradition1                           &  0.62          \\
                                 &                                               & Tradition2                           &  0.62          \\
                                 &                                               & Tradition3                           &  0.51          \\
\multirow{-4}{*}{Tradition}      & \multirow{-4}{*}{0.65}                        & Tradition4                           &  0.56     \\ \hline    

\end{tabular}
}
\end{table*}
    
    As mentioned in Subsection \ref{da}, we calculated Cronbach's alpha to measure the reliability of our survey data and to test the internal consistency of the attributes. Therefore, we tested the 40 attributes within the 10 latent variables (\textit{Benevolence}, \textit{Universalism}, \textit{Self-direction}, \textit{Stimulation}, \textit{Hedonism}, \textit{Achievement}, \textit{Power}, \textit{Security}, \textit{Conformity}, and \textit{Tradition}) using the data from the 193 participants. \autoref{table:cronbach_alpha} shows the results of the measurement reliability test.
    
    We referred the attributes as ``variableN''. For example, the four attributes under the latent variable, \textit{benevolence} are referred as \textit{benevolence1}, \textit{benevolence2}, \textit{benevolence3}, and \textit{benevolence4}). The 10 latent variables (main value categories), their corresponding attributes (value items) with the referred names are shown in \cite{shams_rifat_ara_2022_6370062}. Although there are many controversies with the acceptable range of Cronbach's alpha, the usual interpretation of the coefficient, \(\alpha\) is \cite{ekolu2019reliability}:
    
    \[ \alpha\ < 0.5 = Low \: Reliability \]
    \[ 0.5 < \alpha\ < 0.8 = Acceptable \: Reliability \]
    \[ \alpha\ > 0.8 = High \: Reliability \]
    
    According to the results shown in \autoref{table:cronbach_alpha}, the cronbach's alpha ranged from 0.54 to 0.75 for the ten latent variables. Therefore, all of them were considered acceptable. The obtained reliability measures thus increased the confidence in the contribution of the 40 attributes to the measurement of their respective 10 latent variables. The results of the Cronbach's alpha (if item deleted) did not indicate that any of the items should be removed for \textit{benevolence}, \textit{universalism}, \textit{self-direction}, \textit{achievement}, \textit{security}, \textit{conformity}, and \textit{tradition}. For \textit{stimulation}, \textit{hedonism}, and \textit{power}, Cronbach's alpha could be slightly increased if one item is removed. However, we did not remove any items as the Cronbach's alpha for these variables were already in the acceptable range. These findings provided justification for considering all of the items associated with a particular value.
    

    \subsubsection{Confirmatory Factor Analysis on Individual Value Items}
\begin{figure}[!htbp]
            \centering
            \includegraphics[width=1.03\textwidth]{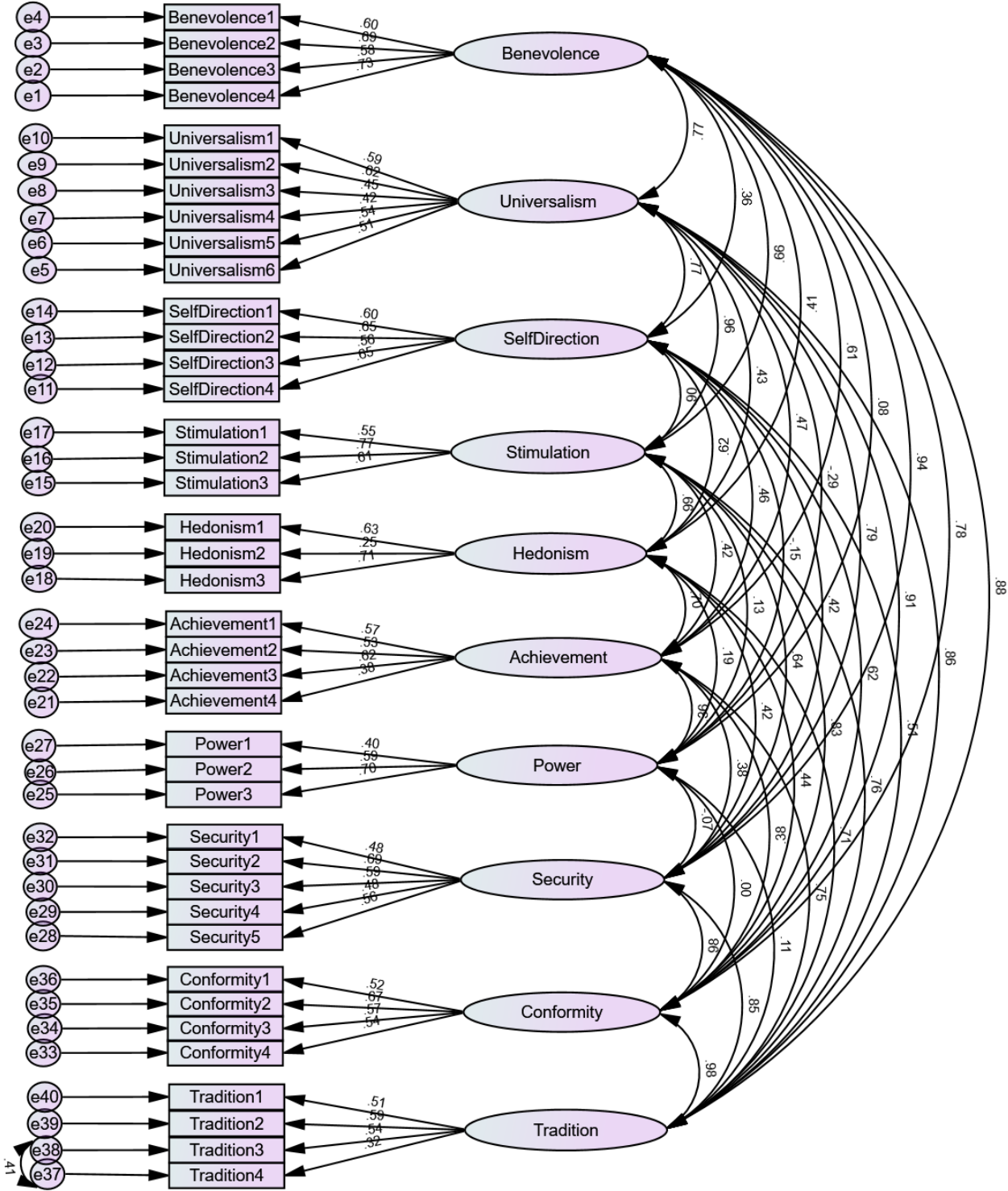}
            \caption{Confirmatory Factor Analysis (CFA) model on 40 value items: Relations of manifest and latent variables}
            \label{fig:ch6_CFA_individual}
        \end{figure}

    As a preparation for exploring the factor structure of Bangladeshi female farmers' values, Confirmatory Factor Analysis (CFA) was conducted on the 40 value items from the survey (PVQ) data using the software, IBM SPSS AMOS 26 (details in step 3 of Subsection \ref{da}). It was conducted to check if the survey data is aligned with Schwartz's theory of basic human values by observing the relations between the manifest variables or observed variables (40 value items) and latent variables or factors. \autoref{fig:ch6_CFA_individual} shows the results of CFA on the 40 value items of the survey data with Bangladeshi female farmers.
        
    The latent and manifest variables are shown in ellipse and rectangle shapes respectively. As mentioned in Subsection \ref{reliability_test}, we refer the value items (manifest variables) as ``variableN'' under a main value category (latent variable) for the convenience of the readers \cite{shams_rifat_ara_2022_6370062}. The unidirectional arrows from the factors to the manifest variables show the direct effects (regressions) of the latent variables onto the observed variables. The weights on the unidirectional arrows refer to the factor loadings or regression coefficients. The curved, bidirectional arrows show the covariances (correlations) among the factors. The term, ``e'' refers to the error terms.
    
    According to \autoref{fig:ch6_CFA_individual}, the cross-loadings are zero, because no indicator (manifest variable) loads on more than one factors. 10 latent variables or factors are extracted from the 40 manifest variables. Four manifest variables of \textit{benevolence} (\textit{benevolence1}, \textit{benevolence2}, \textit{benevolence3}, \textit{benevolence4}) together produce one factor which we also named \textit{benevolence}. Similarly, one factor is extracted from each of the groups of six \textit{universalism} manifest variables, four \textit{self-direction} manifest variables, three \textit{stimulation} manifest variables, three \textit{hedonism} manifest variables, four \textit{achievement} manifest variables, three \textit{power} manifest variables, five \textit{security} manifest variables, four \textit{conformity} manifest variables, and four \textit{tradition} manifest variables. As all the manifest variables of a particular value together extract one factor and it happened for all the 40 value items to extract 10 factors, we argue that the survey data we collected from 193 Bangladeshi female farmers are matched with Schwartz's values theory and good fit for further advanced statistical analysis.
    
    Furthermore, there are also 40 error variances and one error covariance, which means all the error covariances are zero except for the error covariance between \textit{tradition3} and \textit{tradition4}. Another acceptability test of a CFA model is the goodness of fit. Goodness of fit can be evaluated by calculating the p-value, CMIN (chi-square value), and RMSEA (Root Mean Square Error of Approximation) \cite{brown2012confirmatory}. In this CFA model, p-value is .000 (recommended level: $<$0.001 \cite{windle1992revised}), CMIN is 2.782 (recommended level: $<$3.00 \cite{khan2019nexus}), and RMSEA is .085 (recommended level: $<$0.10 \cite{chen2008empirical}). Given that the p-value, CMIN, and RMSEA are in recommended levels, the CFA model is accepted. In addition, all the factor loadings are in acceptable levels ($>$0.32 \cite{kozan2014new}) except for the manifest variable, \textit{hedonism2}. The manifest variables responsible for the lower factor loadings are recommended to remove to develop a better CFA model. However, removing items is not recommended if any value consists of equal or less than three items. As hedonism consists of three items, we did not remove \textit{hedonism2}. However, there are strong correlations among the factors which also make the CFA model acceptable for further analysis.
    

    \subsubsection{Exploratory Factor Analysis on Main Value Categories}
    \label{EFA_Main_Values}
    From the previous step of CFA on value items, ten factors or latent variables (main value categories) were detected. We applied Exploratory Factor Analysis (EFA) on the ten main value categories to explore the possible factors underlying the main value categories. At first, the suitability of the data was verified, and then the factors were extracted.
    
    \textbf{Assessment of the Suitability of the Data.}
    
    \begin{table*}[h]
\centering
\caption{Kaiser-Meyer-Olkin (KMO) and Bartlett's Test}
\label{table:Ch6_KMO}
\resizebox{0.9\textwidth}{!}{%
\renewcommand{\arraystretch}{1.6}
\begin{tabular}{|p{12cm}|p{4cm}|p{4cm}|}

\hline
\multicolumn{2}{|c|}{\textbf{Kaiser-Meyer-Olkin   Measure of Sampling Adequacy}}                      & \multicolumn{1}{|r|}{0.815}   \\ \hline
\multicolumn{1}{|c|}{}                                                         & Approx.   Chi-Square & \multicolumn{1}{|c|}{858.349} \\ \cline{2-3} 
\multicolumn{1}{|c|}{}                                                         & df                   & \multicolumn{1}{|r|}{45}      \\ \cline{2-3} 
\multicolumn{1}{|c|}{\multirow{-3}{*}{\textbf{Bartlett's Test of Sphericity}}} & Sig.                 & \multicolumn{1}{|r|}{0.000}   \\ \hline

\end{tabular}
}
\end{table*}

    The first step of EFA is the assessment of the suitability of the data \cite{hadi2016easy, shrestha2021factor}. Kaiser-Meyer-Olkin (KMO) was used to verify the sampling adequacy \cite{hill2011sequential} and Bartlett’s test of sphericity was used to assess the strength of the inter-correlations among variables \cite{tobias1969brief} using IBM SPSS 26. \autoref{table:Ch6_KMO} shows the Kaiser-Meyer-Olkin (KMO) and Barlett's test results. The KMO presents the suitability of the sampling of this analysis which is 0.815. The KMO score of sampling appropriateness is significant as the KMO should be greater than 0.6 for sampling adequacy \cite{kim2013empirical}. According to the table, Bartlett’s test of sphericity is also acceptable with p-value 0.000 (recommended p-value is $<$0.001 \cite{shiferaw2020validation}).
    
    \textbf{Extraction of the Factors.}
    
    \begin{figure}[htb]
        \centering
        \includegraphics[width=\textwidth]{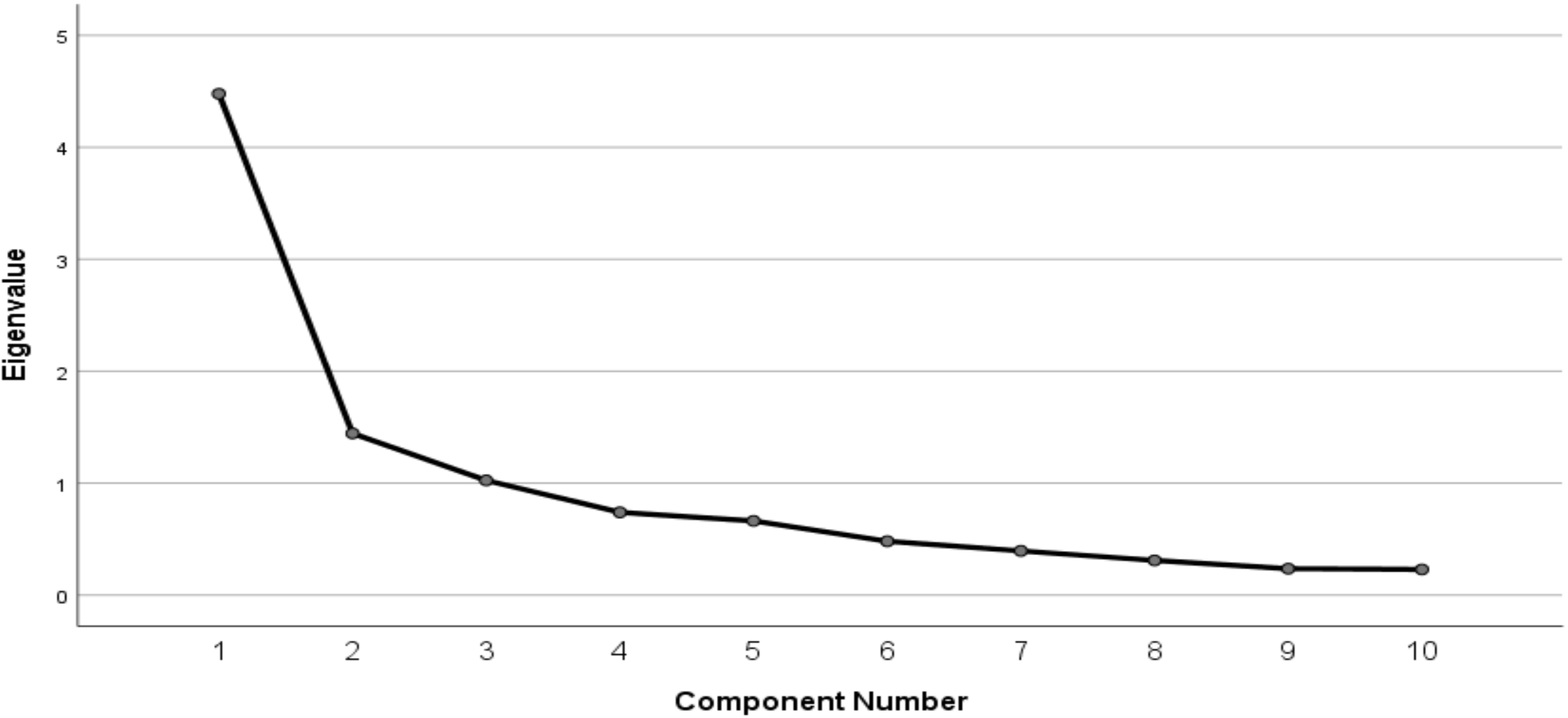}
        \caption{Scree plot}
        \label{fig:EFA_Scree_Plot}
    \end{figure}
    
    Factor extraction is used to identify the least number of factors to represent the interrelationship among the variables \cite{shrestha2021factor}. We conducted EFA with principal component extraction to extract the factors, where the main value categories worked as observed variables. \autoref{table:Ch6_EFA_10_values} shows the total variance. The components whose eigenvalues are greater than one (1.0) are considered in this study as Kaiser’s criterion retains factors with eigenvalue greater than one \cite{suhr2006exploratory}. According to the table, three components or factors are greater than eigenvalues one which explained 69.464\% of the variance of the factors. 50\% is the suggested proportion of the total variance explained by the retained factors \cite{shrestha2021factor}. Therefore, the result is considered acceptable and the factor analysis is considered useful for the variables. The first factor explained 31.648\% of the total variance with eigenvalue 4.478. The second factor explained 21.347\% of the variance with eigenvalue 1.444 and the third factor explained 16.469\% of the total variance with eigenvalue 1.025. \autoref{fig:EFA_Scree_Plot} shows the scree plot, a graphical representation, of the three factors with eigenvalues greater than one. The vertical axis represents the eigenvalue magnitudes and the horizontal axis represents the component number.
    
    \afterpage{
    \clearpage
    \begin{landscape}
    \begin{table*}[htb]
\centering
\caption{Total variance with three components whose eigen value is greater than 1.0}
\label{table:Ch6_EFA_10_values}
\resizebox{1.3\textwidth}{!}{%
{\renewcommand{\arraystretch}{1.4}
\begin{tabular}{|l|l|l|l|l|l|l|l|l|l|}

\hline
\multicolumn{10}{|c|}{\textbf{Total Variance Explained}}                                                                                                                                                                                                                                                                                                                                                                                                                                                                                                                                             \\ \hline
\multicolumn{1}{|l|}{}                            & \multicolumn{3}{c|}{\textbf{Initial Eigenvalues}}                                                                                                            & \multicolumn{3}{c|}{\textbf{Extraction Sums of Squared Loadings}}                                                                                          & \multicolumn{3}{c|}{\textbf{Rotation Sums of Squared Loadings}}                                                                                              \\ \cline{2-10} 
\multicolumn{1}{|l|}{\multirow{-2}{*}{\textbf{Component}}} & \multicolumn{1}{c|}{\textbf{Total}} & \multicolumn{1}{c|}{\textbf{\% of Variance}} & \multicolumn{1}{c|}{\textbf{Cumulative \%}} & \multicolumn{1}{c|}{\textbf{Total}} & \multicolumn{1}{c|}{\textbf{\% of Variance}} & \multicolumn{1}{c|}{\textbf{Cumulative \%}} & \multicolumn{1}{c|}{\textbf{Total}} & \multicolumn{1}{c|}{\textbf{\% of Variance}} & \multicolumn{1}{c|}{\textbf{Cumulative \%}} \\ \hline
\multicolumn{1}{|l|}{1}                           & \multicolumn{1}{r|}{4.478} & \multicolumn{1}{r|}{44.778}        & \multicolumn{1}{r|}{44.778}       & \multicolumn{1}{r|}{4.478} & \multicolumn{1}{r|}{44.778}        & \multicolumn{1}{r|}{44.778}       & \multicolumn{1}{r|}{3.165} & \multicolumn{1}{r|}{31.648}        & \multicolumn{1}{r|}{31.648}       \\ \hline
\multicolumn{1}{|l|}{2}                           & \multicolumn{1}{r|}{1.444} & \multicolumn{1}{r|}{14.441}        & \multicolumn{1}{r|}{59.219}       & \multicolumn{1}{r|}{1.444} & \multicolumn{1}{r|}{14.441}        & \multicolumn{1}{r|}{59.219}       & \multicolumn{1}{r|}{2.135} & \multicolumn{1}{r|}{21.347}        & \multicolumn{1}{r|}{52.995}       \\ \hline
\multicolumn{1}{|l|}{3}                           & \multicolumn{1}{r|}{1.025} & \multicolumn{1}{r|}{10.246}        & \multicolumn{1}{r|}{69.464}       & \multicolumn{1}{r|}{1.025} & \multicolumn{1}{r|}{10.246}        & \multicolumn{1}{r|}{69.464}       & \multicolumn{1}{r|}{1.647} & \multicolumn{1}{r|}{16.469}        & \multicolumn{1}{r|}{69.464}       \\ \hline
\multicolumn{1}{|l|}{4}                           & \multicolumn{1}{r|}{0.739}  & \multicolumn{1}{r|}{7.394}         & \multicolumn{1}{r|}{76.858}       & \multicolumn{1}{l|}{}      & \multicolumn{1}{l|}{}              & \multicolumn{1}{l|}{}             & \multicolumn{1}{l|}{}      & \multicolumn{1}{l|}{}              &                                                           \\ \hline
\multicolumn{1}{|l|}{5}                           & \multicolumn{1}{r|}{0.663}  & \multicolumn{1}{r|}{6.626}         & \multicolumn{1}{r|}{83.484}       & \multicolumn{1}{l|}{}      & \multicolumn{1}{l|}{}              & \multicolumn{1}{l|}{}             & \multicolumn{1}{l|}{}      & \multicolumn{1}{l|}{}              &                                                           \\ \hline
\multicolumn{1}{|l|}{6}                           & \multicolumn{1}{r|}{0.481}  & \multicolumn{1}{r|}{4.811}         & \multicolumn{1}{r|}{88.295}       & \multicolumn{1}{l|}{}      & \multicolumn{1}{l|}{}              & \multicolumn{1}{l|}{}             & \multicolumn{1}{l|}{}      & \multicolumn{1}{l|}{}              &                                                           \\ \hline
\multicolumn{1}{|l|}{7}                           & \multicolumn{1}{r|}{0.395}  & \multicolumn{1}{r|}{3.948}         & \multicolumn{1}{r|}{92.242}       & \multicolumn{1}{l|}{}      & \multicolumn{1}{l|}{}              & \multicolumn{1}{l|}{}             & \multicolumn{1}{l|}{}      & \multicolumn{1}{l|}{}              &                                                           \\ \hline
\multicolumn{1}{|l|}{8}                           & \multicolumn{1}{r|}{0.310}  & \multicolumn{1}{r|}{3.100}         & \multicolumn{1}{r|}{95.342}       & \multicolumn{1}{l|}{}      & \multicolumn{1}{l|}{}              & \multicolumn{1}{l|}{}             & \multicolumn{1}{l|}{}      & \multicolumn{1}{l|}{}              &                                                           \\ \hline
\multicolumn{1}{|l|}{9}                           & \multicolumn{1}{r|}{0.236}  & \multicolumn{1}{r|}{2.364}         & \multicolumn{1}{r|}{97.706}       & \multicolumn{1}{l|}{}      & \multicolumn{1}{l|}{}              & \multicolumn{1}{l|}{}             & \multicolumn{1}{l|}{}      & \multicolumn{1}{l|}{}              &                                                           \\ \hline
\multicolumn{1}{|l|}{10}                          & \multicolumn{1}{r|}{0.229}  & \multicolumn{1}{r|}{2.294}         & \multicolumn{1}{r|}{100.000}      & \multicolumn{1}{l|}{}      & \multicolumn{1}{l|}{}              & \multicolumn{1}{l|}{}             & \multicolumn{1}{l|}{}      & \multicolumn{1}{l|}{}              &                                                           \\ \hline
\multicolumn{10}{|l|}{\textbf{Extraction Method: Principal Component Analysis}}                                                                                                                                                                                                                                                                                                                                                                                                                                                                                                                   \\ \hline

\end{tabular}
}}
\end{table*}

    \end{landscape}
    \clearpage
    }
    
    We also applied varimax rotation with Kaiser normalization as an orthogonal factor rotation technique using IBM SPSS 26 to explore which variables constitute the three factors or components. \autoref{table:Ch6_rotated_component_matrix} shows the rotated component matrix on which the interpretation of factors resulting from the EFA is established. Principal component analysis was undertaken as the extraction method. The rotation converged in five iterations and three factors emerged from the main value categories. The first factor includes five items (main value categories). They are \textit{benevolence}, \textit{security}, \textit{conformity}, \textit{universalism}, and \textit{tradition}. The second factor consists of two items: \textit{self-direction} and \textit{stimulation}. The third factor includes the rest of the three items: \textit{power}, \textit{achievement}, and \textit{hedonism}. According to Kozan et al., the loadings of the items are significant if they are greater than 0.32 \cite{kozan2014new}. In this study, items load strongly on the first factor ranging from 0.573 to 0.858. Similarly, items load significantly on the second factor ranging from 0.727 to 0.859 and on the third factor ranging from 0.634 to 0.790. As all the item loadings on the three factors are extremely strong, it gave us the confidence to apply CFA on the main value categories to verify the factors extracted from EFA.
    
    \begin{table}
\centering
\caption{Rotated component matrix}
\label{table:Ch6_rotated_component_matrix}
\renewcommand{\arraystretch}{1.3}
\begin{tabular}{|lrrr|}
\hline
\multicolumn{4}{|c|}{\textbf{Rotated Component Matrix}} \\ \hline
\multicolumn{1}{|p{7cm}|}{\multirow{2}{*}{\textbf{Manifest Variables}}} & \multicolumn{3}{c|}{\textbf{Component}} \\ \cline{2-4} 
\multicolumn{1}{|p{7cm}|}{} & \multicolumn{1}{c|}{\textbf{1}} & \multicolumn{1}{c|}{\textbf{2}} & \multicolumn{1}{c|}{\textbf{3}} \\ \hline
\multicolumn{1}{|p{7cm}|}{Benevolence} & \multicolumn{1}{r|}{0.858} & \multicolumn{1}{r|}{} &  \\ \hline
\multicolumn{1}{|p{7cm}|}{Security} & \multicolumn{1}{r|}{0.857} & \multicolumn{1}{r|}{} &  \\ \hline
\multicolumn{1}{|p{7cm}|}{Conformity} & \multicolumn{1}{r|}{0.770} & \multicolumn{1}{r|}{} &  \\ \hline
\multicolumn{1}{|p{7cm}|}{Universalism} & \multicolumn{1}{r|}{0.681} & \multicolumn{1}{r|}{} &  \\ \hline
\multicolumn{1}{|p{7cm}|}{Tradition} & \multicolumn{1}{r|}{0.573} & \multicolumn{1}{r|}{} &  \\ \hline
\multicolumn{1}{|p{7cm}|}{Self-Direction} & \multicolumn{1}{r|}{} & \multicolumn{1}{r|}{0.859} &  \\ \hline
\multicolumn{1}{|p{7cm}|}{Stimulation} & \multicolumn{1}{r|}{} & \multicolumn{1}{r|}{0.727} &  \\ \hline
\multicolumn{1}{|p{7cm}|}{Power} & \multicolumn{1}{r|}{} & \multicolumn{1}{r|}{} & 0.790 \\ \hline
\multicolumn{1}{|p{7cm}|}{Achievement} & \multicolumn{1}{r|}{} & \multicolumn{1}{r|}{} & 0.640 \\ \hline
\multicolumn{1}{|p{7cm}|}{Hedonism} & \multicolumn{1}{r|}{} & \multicolumn{1}{r|}{} & 0.634 \\ \hline
\multicolumn{4}{|l|}{Extraction   Method: Principal Component Analysis} \\
\multicolumn{4}{|l|}{Rotation   Method: Varimax with Kaiser Normalization} \\
\multicolumn{4}{|l|}{Rotation   converged in 5 iterations} \\ \hline
\end{tabular}
\end{table}


    \subsubsection{Partial Confirmatory Factor Analysis on Main Value Categories}
    \label{subsubsec:PCFA}
    We conducted Partial Confirmatory Factor Analysis (PCFA), the data reduction technique, to test if the factors extracted from EFA have a solid chance of getting confirmed by CFA. We used IBM SPSS 26 and Excel to conduct PCFA. \autoref{table:Ch6_PCFA} shows the results of PCFA for the four close-fit indexes: Normed Fit Index (NFI), Comparative Fit Index (CFI), Tucker-Lewis Index (TLI), Root Mean Square Error of Approximation (RMSEA) with their corresponding recommended levels.
    
    \begin{table*}
\centering
\caption{The results of the close-fit indexes of Partial Confirmatory Factor Analysis (PCFA)}
\label{table:Ch6_PCFA}
\resizebox{\textwidth}{!}{%
\renewcommand{\arraystretch}{1.6}
\begin{tabular}{|p{6cm}|p{4cm}|p{6cm}|}

\hline
\multicolumn{1}{|c|}{\textbf{Close-fit   Indexes}} & \multicolumn{1}{c|}{\textbf{Results}} & \textbf{Recommended Levels} \\ \hline
\textbf{NFI}                                       & 0.923                                 & $>$0.90 \cite{cann2011assessing}                      \\ \hline
\textbf{CFI}                                       & 0.941                                 & $>$0.90 \cite{cann2011assessing}                       \\ \hline
\textbf{TLI}                                       & 0.851                                 & $>$0.89 \cite{ryberg2020measuring}                       \\ \hline
\textbf{RMSEA}                                     & 0.118                                 & $<$0.18 \cite{cann2011assessing}                       \\ \hline

\end{tabular}
}
\end{table*}
    
    The table shows that the NFI, CFI, and RMSEA are at acceptable levels. However, TLI is slightly lower than the recommended level. As the other three close-fit indexes are in the accepted range and TLI is close to the recommended level, we argue that the factors extracted from EFA have a high chance of getting confirmed by CFA. Therefore, conducting CFA is justified.
    

    \subsubsection{Confirmatory Factor Analysis on Main Value Categories}
\begin{figure}[htb]
        \centering
        \includegraphics[width=0.73\textwidth]{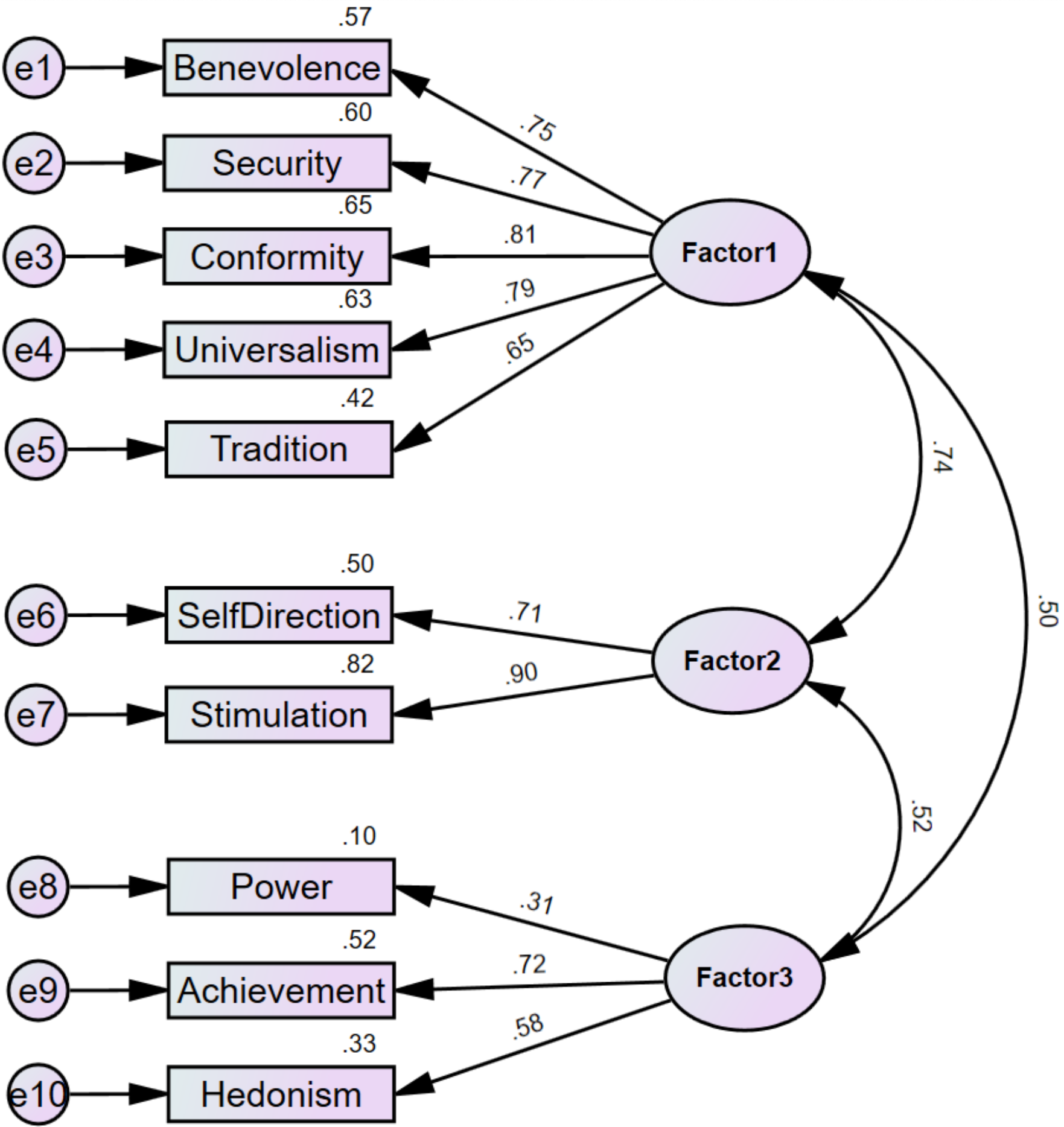}
        \caption{Confirmatory Factor Analysis (CFA) model on 10 main value categories: Extraction of factors}
        \label{fig:ch6_CFA_10_values}
    \end{figure}

    We conducted Confirmatory Factor Analysis (CFA) on the ten main value categories using IBM SPSS AMOS 26 to verify the relations between the 10 main value categories (manifest variables) and the factors or components (latent variables) derived from the Exploratory Factor Analysis (EFA). \autoref{fig:ch6_CFA_10_values} shows the results of CFA on the main value categories of the survey conducted with Bangladeshi female farmers. The factors and the main value categories are shown in the ellipse and rectangle shapes respectively. The model extracted three factors. The first factor (Factor1) consists of five manifest variables (main value categories). They are, \textit{benevolence}, \textit{security}, \textit{conformity}, \textit{universalism}, and \textit{tradition}. The second factor (Factor2) includes two variables: \textit{self-direction} and \textit{stimulation}. The third factor (Factor3) consists of three variables: \textit{power}, \textit{achievement}, and \textit{hedonism}. The result is exactly similar to the result of EFA (see \autoref{table:Ch6_rotated_component_matrix}) discussed in Subsection \ref{EFA_Main_Values}.
    
    As shown in \autoref{fig:ch6_CFA_10_values}, no variable loads on more than one factors. Hence, the cross-loadings of CFA are zero. Furthermore, there are ten error variances with zero error covariances. Therefore, the CFA model is considered acceptable. In addition, all the factor loadings in this model are in the acceptable range of equal or above 0.32 \cite{kozan2014new} except for the factor loading of \textit{power}. As the factor loading for \textit{power} is 0.31 which is close to 0.32, \textit{power} was not removed from the CFA model. However, the inter-correlations among the factors (latent variables) should be below 0.85 to confirm the CFA model \cite{awang2015evaluation}. As shown in \autoref{fig:ch6_CFA_10_values}, the model also supports this condition, where all the correlations among factors are below 0.85. They are 0.50 between Factor1 and Factor3, 0.74 between Factor1 and Factor2, and 0.52 between Factor2 and Factor3. Further goodness of fit evaluation was conducted in Subsection \ref{subsubsec:PCFA} which also enhances the acceptability of this model.
    

\subsection{RQ2: Influence of Demographics on the Values of Bangladeshi Female Farmers}
\label{Results_RQ2}
As described in Section \ref{subsec:survey}, six demographic data (area, age, education level, household income, number of educated family members, highest level of education among family members) of Bangladeshi female farmers were collected during the survey. One of our paper discussed how value priorities of Bangladeshi female farmers differ demographically through descriptive statistical analysis \cite{shams2021measuring}. This section presents the significance of demographics (independent variables) on the main value categories of Bangladeshi female farmers (dependent variables). For this purpose, we conducted an advanced statistical method, multiple linear regression method, using IBM SPSS 26. \autoref{tab:Ch6_Regression} shows the result of the influence of demographics on the values of Bangladeshi female farmers.
    
    \afterpage{
    \clearpage
    \begin{landscape}
    
    \begin{table*}[htb]
\centering
\caption{Results of regression analysis: Significance of demographics on values of Bangladeshi female farmers}
\label{tab:Ch6_Regression}
\resizebox{1.4\textwidth}{!}{%
{\renewcommand{\arraystretch}{1.6}
\begin{tabular}{|l|r|r|r|r|r|r|r|r|r|r|}

\hline
\multicolumn{1}{|c|}{\multirow{2}{*}{\textbf{Demographics}}}        & \multicolumn{10}{c|}{\textbf{Significance on values (p-value) }}                                                                                                                                                          \\ 
\cline{2-11}
                                              & \textbf{Benevolence} & \textbf{Universalism} & \textbf{Self-direction} & \textbf{Stimulation} & \textbf{Hedonism} & \textbf{Achievement} & \textbf{Power} & \textbf{Security} & \textbf{Conformity} & \textbf{Tradition}  \\ 
\hline
\textbf{Age}                                  & 0.145                & \textbf{0.083*}                & 0.550                   & 0.979                & 0.239             & 0.243                & 0.280          & \textbf{0.037**}           & 0.136               & 0.110               \\ 
\hline
\textbf{Education level}                      & 0.414                & 0.352                 & 0.304                   & 0.784                & \textbf{0.083*}            & \textbf{0.082*}               & 0.874          & 0.964             & 0.928               & 0.598               \\ 
\hline
\textbf{No. of educated family member}         & 0.172                & 0.280                 & \textbf{0.089*}                  & 0.217                & 0.835             & 0.217                & 0.461          & 0.900             & 0.235               & 0.113               \\ 
\hline
\textbf{Area}                                 & \textbf{0.063*}               & 0.274                 & 0.831                   & 0.138                & \textbf{0.000***}          & \textbf{0.000***}             & 0.128          & 0.378             & 0.759               & \textbf{0.001***}            \\ 
\hline
\textbf{Household income}                     & \textbf{0.087*}               & 0.100                 & 0.250                   & 0.138                & 0.868             & 0.777                & \textbf{0.001***}       & \textbf{0.004***}          & \textbf{0.094*}              & 0.205               \\ 
\hline
\textbf{Highest level of education in family} & 0.852                & 0.731                 & 0.666                   & 0.947                & 0.663             & 0.524                & 0.775          & 0.547             & 0.871               & 0.250               \\

\hline
\multicolumn{11}{l}{}\\
\multicolumn{11}{l}{\textbf{*** Strong significance $(p-value \textless{} 0.01)$ }}                                                                                                                                                                                          \\
\multicolumn{11}{l}{\textbf{** Moderate significance $(0.01 \le p-value \textless{} 0.05)$ }}                                                                                                                                                                         \\
\multicolumn{11}{l}{\textbf{* Weak significance $(0.05 \le p-value \textless{} 0.10)$ }}                                                                                                                                                                              \\

\end{tabular}%
} \quad
}
\end{table*}
    \begin{table*}[htb]
\centering
\caption{Results of regression analysis: Presence of multicollinearity}
\label{tab:Ch6_Regression_VIF}
\resizebox{1.4\textwidth}{!}{%
{\renewcommand{\arraystretch}{1.6}
\begin{tabular}{|l|r|r|r|r|r|r|r|r|r|r|}

\hline
\multicolumn{1}{|c|}{\multirow{2}{*}{\textbf{Demographics}}}        & \multicolumn{10}{c|}{\textbf{Variance Inflation Factor (VIF)}}                                                                                                                                                          \\ 
\cline{2-11}
                                              & \textbf{Benevolence} & \textbf{Universalism} & \textbf{Self-direction} & \textbf{Stimulation} & \textbf{Hedonism} & \textbf{Achievement} & \textbf{Power} & \textbf{Security} & \textbf{Conformity} & \textbf{Tradition}  \\ 
\hline
\textbf{Age}                                  & 1.326                & 1.326                & 1.326                   & 1.326                & 1.326             & 1.326                & 1.326          & 1.326           & 1.326               & 1.326               \\ 
\hline
\textbf{Education level}                      & 1.233                & 1.233                 & 1.233                   & 1.233                & 1.233            & 1.233               & 1.233          & 1.233             & 1.233               & 1.233               \\ 
\hline
\textbf{No. of educated family member}         & 2.031                & 2.031                 & 2.031                  & 2.031                & 2.031             & 2.031                & 2.031          & 2.031             & 2.031               & 2.031               \\ 
\hline
\textbf{Area}                                 & 1.116               & 1.116                 & 1.116                   & 1.116                & 1.116          & 1.116             & 1.116          & 1.116             & 1.116               & 1.116            \\ 
\hline
\textbf{Household income}                     & 1.407               & 1.407                 & 1.407                   & 1.407                & 1.407             & 1.407                & 1.407       & 1.407          & 1.407              & 1.407               \\ 
\hline
\textbf{Highest level of education in family} & 1.957                & 1.957                 & 1.957                   & 1.957                & 1.957             & 1.957                & 1.957          & 1.957             & 1.957               & 1.957               \\

\hline
\multicolumn{11}{l}{}\\
\multicolumn{11}{l}{\textbf{High presence of multicollinearity, if VIF $>$ 10}}                                                                                                      

\end{tabular}%
} \quad
}
\end{table*}
    \begin{table*}[htb]
\centering
\caption{Results of regression analysis: R Square}
\label{tab:Ch6_Regression_R_Square}
\resizebox{1.4\textwidth}{!}{%
{\renewcommand{\arraystretch}{1.6}
\begin{tabular}{|l|r|r|r|r|r|r|r|r|r|r|}

\hline
\multicolumn{1}{|c|}{\multirow{2}{*}{\textbf{Predictors}}}        & \multicolumn{10}{c|}{\textbf{R Square}}                                                                                                                                                          \\ 
\cline{2-11}
                                              & \textbf{Benevolence} & \textbf{Universalism} & \textbf{Self-direction} & \textbf{Stimulation} & \textbf{Hedonism} & \textbf{Achievement} & \textbf{Power} & \textbf{Security} & \textbf{Conformity} & \textbf{Tradition}  \\ 
\hline
\textbf{Demographics}                                  & 0.053                & 0.041                & 0.032                   & 0.032                & 0.118             & 0.169                & 0.086          & 0.087           & 0.041               & 0.110               \\ 
\hline

\end{tabular}%
} \quad
}
\end{table*}

    \end{landscape}
    \clearpage
    }
    
There are strong significance of independent variables on dependent variables if the p-value (significance) is less than 0.01, moderate significance if the p-value is greater than or equal to 0.01 and less than 0.05, weak significance if the p-value is greater than or equal to 0.05 and less than 0.10 \cite{zanutto2006comparison}. There are no influences of independent variables on dependent variables if the p-value is greater than or equal to 0.10 \cite{zanutto2006comparison}. According to \autoref{tab:Ch6_Regression}, \textit{hedonism}, \textit{achievement}, and \textit{tradition} have strong correlations with area. The p-values are 0.000 for \textit{hedonism} and \textit{achievement}, while 0.001 for \textit{tradition}. Similarly, \textit{power} and \textit{security} have strong correlations with household income (p-values are 0.001 and 0.004 respectively). There are moderate significance of age on \textit{security}, the p-value is 0.037. However, all the demographics except ``highest level of education in family'' have weak significance on some of the values. For example, age has weak influence on \textit{universalism} (p-value is 0.083), while education level has weak correlations with \textit{hedonism} and \textit{achievement} (p-values are 0.083 and 0.082 respectively). Similarly, there are weak significance of the number of educated family members on \textit{self-direction} (p-value is 0.089), area on \textit{benevolence} (p-value is 0.063), and household income on \textit{benevolence} (p-value is 0.087) and \textit{conformity} (p-value is 0.094). However, the demographics have no significant influences on the rest of the values.
    
\autoref{tab:Ch6_Regression_VIF} shows the presence of multicollinearity in the regression analysis through the Variance Inflation Factor (VIF). Multicollinearity investigates the presence of linear relations among independent variables \cite{uddin2021did}. In other words, multicollinearity occurs when there are strong correlations among two or more independent variables. There is a high presence of multicollinearity in regression analysis if the VIF value is greater than ten (10) \cite{uddin2021did}. As shown in \autoref{tab:Ch6_Regression_VIF}, all the VIF values are less than ten. Therefore, multicollinearity is not present in the regression analysis.
    
\autoref{tab:Ch6_Regression_R_Square} shows the results of $R^2$ in the regression analysis. The demographics act as the most significant predictors for \textit{achievement}, as they explain 16.9\% (0.169) of the variance. The scores of $R^2$ are also high for \textit{hedonism} and \textit{tradition}, 11.8\% (0.118) and 11\% (0.110) respectively, resulting the demographics are also significant for these two values. The demographics work as the least significant predictors for \textit{self-direction} and \textit{stimulation}, as they explain 3.2\% (0.032) of the variance for both.
\section{Discussion and Implications}
\label{sec:discussion}

\subsection{Analysis of the Results}
The results of RQ1 (see Section \ref{Results_RQ1}) identified three underlying factors of Bangladeshi female farmers' values. In other words, three groups of Bangladeshi female farmers emerged according to their value preferences. The first group can be considered as benevolent, conscious, respectful, universalist, and conservative. This is because, the first factor consists of five values: \textit{benevolence}, \textit{security}, \textit{conformity}, \textit{universalism}, and \textit{tradition}. Similarly, the second group can be considered as self-directed and adventurous, as Factor2 comprises of two values, \textit{self-direction} and \textit{stimulation}. The third group of Bangladeshi female farmers can be considered as hedonist, aspiring, and powerful due to the three values (\textit{power}, \textit{achievement}, and \textit{hedonism}) which are included in Factor3.

After comparing the preferred values of these three groups with Schwartz's theory of basic values (see \autoref{fig:SchwartzTheory}), we argue that the results of RQ1 are in line with Schwartz's theory. For example, \textit{universalism}, \textit{benevolence}, \textit{conformity}, \textit{tradition}, and \textit{security} are the important values for the participants of the first group. All these values are located close to each other in Schwartz's theory as well. The participants of the second group preferred \textit{self-direction} and \textit{stimulation}, which are also located next to each other in Schwartz's theory. The participants of the last group consists of the values, \textit{power}, \textit{achievement}, and \textit{hedonism}, which are also placed together in Schwartz's theory. According to Schwartz, values located close to each other are congruent and those further apart are opposite in nature \cite{schwartz1992universals, schwartz2012overview}. Therefore, it verified the accuracy of the results of the factor analysis, where all the three groups included values which are congruent in nature.

The results of RQ1 can also be compared with the results of one of our papers that measured Bangladeshi female farmers' value priorities \cite{shams2021measuring}. For example, Bangladeshi female farmers have positive priorities for all the five values which are grouped together in the first factor. On the other hand, \textit{power} and \textit{hedonism} are the least important values of Bangladeshi female farmers. Also, these two values together develop the third factor. However, the second factor consists of \textit{self-direction} and \textit{stimulation}, whereas \textit{self-direction} is positive and \textit{stimulation} is negative priority for Bangladeshi female farmers.

    \begin{tcolorbox}[colback=gray!5!white,colframe=gray!75!black,title=Analysis of the Results]
        \justify
        Bangladeshi female farmers can be divided into three groups according to their values. The participants of the first group are benevolent, conscious, respectful, universalist, and conservative. The second group participants are self-directed and adventurous. The third group participants are hedonist, aspiring, and powerful.
    \end{tcolorbox}


\subsection{Implications for Apps Development}
One of our papers identified the overall value priorities of Bangladeshi female farmers \cite{shams2021measuring}. If the developers develop agriculture apps based on the most preferred values shown in this paper \cite{shams2021measuring}, majority of the Bangladeshi female farmers' values should be addressed in the apps. However, there is also a chance that a small group of end-users' values might be ignored in those apps if they have a different value priorities. To avoid this issue, this study investigated if there are any groups of Bangladeshi female farmers who have the similar preferred values. As this study identified three groups of Bangladeshi female farmers according to their value preferences, agriculture apps should be different for these three groups. We recommend applying three different apps design strategies for these three groups of end-users of agriculture apps. In particular, we recommend apps developers to develop agriculture apps for Bangladeshi female farmers considering the specific values depending on which group(s) are the target end-users. For example, if the target users are benevolent, conscious, respectful, universalist, and conservative Bangladeshi female farmers (group 1), the apps should reflect \textit{benevolence}, \textit{security}, \textit{conformity}, \textit{universalism}, and \textit{tradition}. On the other hand, \textit{self-direction}, \textit{stimulation}, \textit{power}, \textit{achievement}, and \textit{hedonism} should not be reflected in those apps, because these values are not preferred by the target end-users. Similarly, the agriculture apps for Bangladeshi female farmers should be different depending on the target end-users' demographics as well. Developers need to be conscious about addressing the values that have strong correlations with demographics during apps development for different demographic groups. Therefore, we also recommend further research to propose strategies for apps development for different demographic groups.

    \begin{tcolorbox}[colback=gray!5!white,colframe=gray!75!black,title=Implication 1 for Apps Development]
        \justify
        Bangladeshi agriculture apps developers are recommended to apply a particular set of apps design strategies and/or approach for a specific group of the target end-users according to their value preferences.
    \end{tcolorbox}
    
    The findings of this study encourage developers to pay attention to end-users' values during apps development to promote the use of the apps. However, it can be argued that embedding end-users' existing values into an app does not always have positive impacts on end-users' lives. Designing apps that reinforce existing value sets might cause negative impacts on human lives. This is because if a society arguably has negative values, addressing those values in apps might amplify the negativity. For example, a subjugated population might have values that reflect subjugation. Therefore, we recommend further research to understand when reinforcing existing value sets in apps design might have positive impacts and when negative. As people ubiquitously use apps in their daily activities nowadays, this is an opportunity to use apps to discourage the negative values of the end-users and help free them from those negative values. Therefore, we recommend to investigate the extent to which values-based design of mobile apps should embed existing set of end-users' values and what attempts can be taken to nudge end-users towards an alternative set of values.
    
    \begin{tcolorbox}[colback=gray!5!white,colframe=gray!75!black,title=Implication 2 for Apps Development]
        \justify
        Further research is recommended to explore the extent to which values-based design of mobile apps should embed existing set of end-users' values and what attempts can be taken to nudge end-users towards an alternative set of values.
    \end{tcolorbox}
     
\section{Threats to Validity}
\label{sec:ttv}

This section discusses the possible threats arising from this research according to the four validation criteria: credibility, confirmability, dependability, and transferability \cite{cruzes2011recommended}.

    \subsection{Credibility}
    A potential threat to the credibility of this research could arise from participants' selection approach. To mitigate this threat, we requested help from the senior employees of Oxfam Bangladesh. They have several years of experience working with Bangladeshi female farmers. After knowing the objectives of this research and participants' selection criteria, they introduced us to the right participants.
    

    \subsection{Confirmability}
    \label{ttv_confirmability}
    A potential threat to confirmability could arise from the absence of data-source triangulation and the validity of the results. We accept that data-source triangulation could verify the results and increase the confirmability of this research. However, we mitigated this threat by using a universal value measurement instrument which was used to examine the cross-cultural validity of Schwartz's values theory \cite{schwartz2001extending}. It was used in several countries with different cultural settings to investigate human values. Furthermore, the large number of participants in this survey also increased the plausibility of our findings.
    

    \subsection{Dependability}
    A potential threat to the dependability of this research could arise from the lack of understanding of human values. This is because of the ill-defined, ambiguous, and implicit nature of human values \cite{perera2020study} and the absence of definitions of human values from mobile apps/software engineering perspectives. However, no questions in PVQ contained the term ``values'' directly, which minimized the chance of misunderstanding this term.
    

    \subsection{Transferability}
    A potential threat to the transferability of this research could arise due to the focus on a specific group of end-users, Bangladeshi female farmers. As different groups of end-users might have different values when using different apps, it can be argued that the results of this research cannot be generalized for all the end-users of mobile apps. However, we believe the results can be used for the users and apps of other developing countries like Bangladesh. Furthermore, the methodology used in this research can be replicated for the users and apps in different cultural settings in other countries.
\section{Conclusions and Future Work}
\label{sec:conclusion}

\subsection{Conclusions}
Exploratory Factor Analysis (EFA) and Confirmatory Factor Analysis (CFA) were conducted to explore the possible underlying factor structure of Bangladeshi female farmers' values (RQ1). CFA on the 40 value items of the survey with 193 Bangladeshi female farmers shows that the survey data is aligned with Schwartz's theory and a good fit for advanced statistical analysis. EFA was also conducted on the ten main value categories which identified three (3) underlying factors of Bangladeshi female farmers' values. The first factor comprises of five values: \textit{benevolence}, \textit{security}, \textit{conformity}, \textit{universalism}, and \textit{tradition}. The second factor consists of two values: \textit{self-direction} and \textit{stimulation}. The third factor includes the rest of the three values: \textit{power}, \textit{achievement}, and \textit{hedonism}. CFA was again conducted on the main value categories which confirmed the findings of the EFA by developing a CFA model. Considering the three factors, Bangladeshi female farmers can be divided into three groups according to their values. The participants of the first group are benevolent, conscious, respectful, universalist, and conservative. The second group participants are self-directed and adventurous. The third group participants are hedonist, aspiring, and powerful.

A multiple linear regression method was also applied on the survey data to explore the influences of demographics on Bangladeshi female farmers' values (RQ2). The results identified strong influences of area on three values: \textit{hedonism}, \textit{achievement}, and \textit{tradition}. There are also strong influences of household income on \textit{power} and \textit{security}. Moderate influence of age was identified on \textit{security}. However, age has weak influence on \textit{universalism}; education level on \textit{hedonism} and \textit{achievement}; number of educated family members on \textit{self-direction}; area on \textit{benevolence}; household income on \textit{benevolence} and \textit{conformity}.

This research creates awareness among software engineering researchers and software applications developers to consider human values in mobile apps. The findings of this research provide implications for software engineering research and practices on how to measure the factor structure of end-users' values who are vulnerable groups of women, which of their values should be addressed during apps development for different value groups and different demographic groups of end-users.


\subsection{Future Work}
\label{future_work}
The replication of this research with larger samples from different cultural settings could be a significant future work to observe if there are any cultural differences regarding the values of end-users of mobile apps. For example, this research could be replicated with Bangladeshi male farmers to determine the extent to which the values are gender-specific. Empirical data from other end-users of apps from different cultural settings could also be collected to ensure the transferability of this research. In addition, it would be interesting to conduct a similar empirical study of the agriculture mobile apps of other developing countries. It would provide an opportunity to compare the values of the end-users.

As we identified three different groups of Bangladeshi female farmers according to their values, the agriculture apps developed for them should also be different. Although one of our paper proposed strategies to address Bangladeshi female farmers' values in agriculture apps \cite{shams2021human}, we recommend developing different apps design strategies for different groups of Bangladeshi female farmers. Therefore, there is room for research to develop agriculture apps for Bangladeshi female farmers considering the specific values depending on which group(s) are the target end-users.

A potential research direction could focus on whether embedding end-users' existing values into an app always has positive impacts on end-users' lives. If the users arguably have negative values (e.g., subjugation, exercising power), addressing those values in apps might amplify the negativity. Therefore, extensive research in social science, psychology, geopolitics, and behavioural science is required to understand when reinforcing existing value sets in apps design might have positive impacts and when negative. If the users have negative values, apps should actively try to discourage those. Therefore, a potential future work could investigate the extent to which apps should embed existing end-users' values and what attempts can be taken to nudge end-users towards an alternative set of values.




\Urlmuskip=0mu plus 1mu\relax
\bibliographystyle{elsarticle-num}
\bibliography{References.bib}





\end{document}